\input epsf
\input rotate                
\documentstyle[12pt,epsf,colordvi,rotate]{article}
\newlength{\reducedwidth}
\newcommand{\beq}{\begin{equation}}
\newcommand{\eeq}{\end{equation}}
\newcommand{\beqs}{\begin{eqnarray}}
\newcommand{\eeqs}{\end{eqnarray}}
\newcommand{\lsim}{\mathrel{\raisebox{-.6ex}{$\stackrel{\textstyle<}{\sim}$}}}
\newcommand{\gsim}{\mathrel{\raisebox{-.6ex}{$\stackrel{\textstyle>}{\sim}$}}}

\title{Oscillation Measurements with Upgraded Conventional Neutrino Beams}

\font\eightit=cmti10
\def\r#1{\ignorespaces $^{#1}$}

\begin{document}
\maketitle  

%\huge
\begin{center}
%{\bf Oscillation Measurements with Upgraded Conventional Neutrino Beams}

%\hfilneg
\begin{sloppypar}
\noindent
\large

V. Barger, \r{1}
R. Bernstein, \r{4}
A. Bueno, \r{2}
M. Campanelli, \r{2}
D. Casper, \r{3}\\
F. DeJongh, \r{4} 
S.Geer, \r{4} 
M. Goodman, \r{5} 
D.A. Harris, \r{4} 
K.S. McFarland, \r{6}\\
N. Mokhov, \r{4}
J. Morfin,\r{4}
J. Nelson,\r{7}
F. Pietropaolo, \r{8}
R. Raja,\r{4} 
J. Rico, \r{2}\\
A. Rubbia, \r{2}
H. Schellman, \r{9}
R. Shrock, \r{10}
P. Spentzouris,\r{4}
R. Stefanski, \r{4}\\
L. Wai, \r{11}
K. Whisnant \r{12}

\end{sloppypar}

\normalsize
\vskip .4in

\r{1} {\eightit University of Wisconsin, Madison, WI 53706} \\
\r{2} {\eightit Institut f\"ur Teilchenphysik, ETHZ, CH-8093, Z\"urich, Switzerland} \\
\r{3} {\eightit University of California Irvine, Irvine, CA 92697} \\
\r{4} {\eightit Fermi National Accelerator Laboratory, Batavia, IL 60510} \\
\r{5} {\eightit Argonne National Laboratory, Argonne, IL 60439} \\
\r{6} {\eightit University of Rochester, Rochester, NY 14627} \\
\r{7} {\eightit University of Minnesota, Minneapolis, MN 55455} \\
\r{8} {\eightit University of Padova, Padova, Italy} \\
\r{9} {\eightit Northwestern University, Evanston, IL 60208} \\
\r{10} {\eightit State University of New York Stony Brook, Stony Brook, NY 11794}\\
\r{11} {\eightit Stanford University, Stanford, CA} \\
\r{12} {\eightit Iowa State University, Ames, IA  50011} \\
\end{center}
                                                          
\newpage

\abstract{
We consider the $\nu_\mu \to \nu_e$ oscillation measurements 
that would be possible at upgraded 1~GeV and multi--GeV 
conventional neutrino sources driven 
by future megawatt--scale proton drivers. If these 
neutrino superbeams are used 
together with detectors that are an order of magnitude larger 
than those presently foreseen, we find that the 
sensitivity to $\nu_\mu \to \nu_e$ 
oscillations can be improved by an order of magnitude beyond the next 
generation of accelerator based experiments. In addition, over a limited 
region of parameter space, the neutrino mass hierarchy can be determined 
with a multi--GeV long baseline beam. If the Large Mixing Angle 
MSW solution correctly 
describes the solar neutrino deficit, there is a small corner of allowed 
parameter space in which maximal CP--violation in the lepton sector might 
be observable at a 1~GeV medium baseline experiment. 
Superbeams with massive detectors would therefore provide a useful tool 
en route to a neutrino factory, which would permit a further order of 
magnitude improvement in sensitivity, together with a more comprehensive 
check of CP--violation and the oscillation framework.
} 

\newpage

\section{Prologue}

We have recently completed a six--month study of 
the prospective physics program at a neutrino factory, 
evaluated as a function of the stored muon energy (up to 
50~GeV) and the number of useful muon decays per year 
(in the range from $10^{19}$ to $10^{21}$). 
The basic conclusions presented in 
our report~\cite{report} were that: (1) There is a compelling 
physics case for a neutrino factory~\cite{nufact} with a muon beam energy 
of about 20~GeV or greater, and (2) The neutrino factory should 
provide at least O($10^{19}$)
useful decays per year initially, and ultimately at least 
O($10^{20}$) decays per year. The  
oscillation physics that can be pursued using initial 
electron--neutrino ($\nu_e$) and electron--antineutrino 
($\overline{\nu}_e$) beams provides the primary motivation 
for a neutrino factory. In particular we found that 
with $2 \times 10^{20}$ decays per year, after a few years 
of running:
\begin{description}
\item{(i)} 
A $\nu_e \rightarrow \nu_\mu$ oscillation signal could 
be observed, and the associated amplitude parameter 
$\sin^2 2\theta_{13}$ measured, for oscillation amplitudes 
approaching $10^{-4}$, three orders of magnitude below the 
currently excluded region and two orders of magnitude below 
the region expected to be probed by the next generation of 
long--baseline accelerator based experiments. 
\item{(ii)}
Once a $\nu_e \rightarrow \nu_\mu$ signal has been established 
in a 
long baseline experiment, matter effects can be exploited to 
determine the sign of the difference between the squares 
of the neutrino mass eigenstates $\delta m^2_{32}$, and 
hence determine the neutrino mass hierarchy.
\item{(iii)}
If the large mixing angle MSW solution describes the solar neutrino 
deficit, and if $\sin^2 2\theta_{13}$ is not less than one to two 
orders of magnitude below the currently excluded region, 
a comparison of $\nu_e \rightarrow \nu_\mu$ and 
$\overline{\nu}_e \rightarrow \overline{\nu}_\mu$ oscillation 
probabilities would enable the measurement of (or stringent limits 
on) CP--violation in the lepton sector.
\item{(iv)}
Measurements of, or stringent limits on, all of the observable 
$\nu_e \rightarrow \nu_X$ oscillation modes together with 
the observable $\nu_\mu \rightarrow \nu_X$ modes, would enable a 
comprehensive test of the assumed oscillation framework.
\end{description}

In parallel with our neutrino factory physics study, a companion 
design study~\cite{norbert} was conducted to determine the feasibility 
of constructing 
a neutrino factory, and to identify the associated R\&D issues. The 
design study concluded that a neutrino factory with the 
desired parameters was indeed feasible, although it would require 
a vigorous and well supported R\&D program. Recognizing that the R\&D 
would take some time, and that a neutrino factory would require a 
very intense (megawatt--scale) proton driver, it is reasonable to consider 
the neutrino oscillation physics program that could be conducted 
using a MW--scale proton driver en route to a neutrino factory. 
In our neutrino factory physics study report we recommended that 
an additional study of the oscillation physics 
potential at these ``neutrino superbeams'' be undertaken. 

The present document, which can be considered as an addendum to our 
initial report, presents results from a study of oscillation physics 
at 1~GeV and multi--GeV neutrino superbeams.

\section{Introduction}

In this report we consider the oscillation physics capabilities 
of neutrino ``superbeams'', which we define as conventional 
neutrino beams produced using megawatt--scale high--energy proton drivers. 
Examples of appropriate proton drivers are 
(i) the proposed 0.77~MW 50~GeV proton synchrotron 
at the Japan Hadron Facility (JHF)~\cite{jhfloi}, 
(ii) a 4~MW upgraded version of the JHF, 
(iii) a new $\sim 1$~MW 16~GeV proton driver that would replace the existing 
8~GeV Booster at Fermilab, or 
(iv) a fourfold intensity upgrade of the 120~GeV Fermilab Main Injector (MI) 
beam (to 1.6~MW) that would become possible once the upgraded (16~GeV) 
Booster was operational.  
The 4~MW 50~GeV JHF and the 16~GeV upgraded Fermilab Booster, are 
both suitable proton drivers for a neutrino factory. Hence a neutrino 
superbeam might provide a neutrino physics program en route to a neutrino 
factory. 

The next generation of accelerator based long--baseline 
neutrino oscillation experiments 
are expected to confirm the $\nu_\mu \rightarrow \nu_\tau$ 
oscillation interpretation of the atmospheric 
muon--neutrino deficit, and begin to measure the associated oscillation 
parameters with modest statistical precision. To make a significant 
improvement in oscillation measurements beyond the next generation of 
experiments will require a significant increase in signal statistics. 
A factor of a few increased neutrino beam flux will not be sufficient 
unless there is also a substantial increase in detector mass. We will 
therefore assume that superbeam experiments will use detectors an order of 
magnitude larger than those currently under construction. Hence, a superbeam 
experiment would yield data samples with a statistical sensitivity 
a factor of at least $\sqrt{40}$ better than expected for 
K2K~\cite{k2k}, MINOS~\cite{minostdr}, OPERA~\cite{opera}, 
and ICARUS~\cite{icarus}. Clearly, this would enable significant progress in 
pinning down $\nu_\mu \rightarrow \nu_\tau$ oscillations, for example. 
However, the neutrino oscillation physics program at a superbeam would 
have to justify the substantial investment associated with the detector.
For example, a detector costing ten times the MINOS detector would be 
of order \$300M. 
Hence, more precise measurements of the quantities already 
measured by the next generation experiments may be insufficient motivation 
for a superbeam. 

The primary motivation for a neutrino superbeam is likely to be the 
search for, and measurement of, $\nu_\mu \rightarrow \nu_e$ oscillations. 
The observation of this mode would enable the associated amplitude parameter 
$\sin^2 2\theta_{13}$ to be determined, and open the way for the determination 
of the sign of the differences between the squares of the neutrino mass 
eigenstates $\delta m^2_{32}$, and hence the determination of the 
neutrino mass hierarchy. A 
comparison between the oscillation probabilities for 
$\nu_\mu \rightarrow \nu_e$ and 
$\overline{\nu}_\mu \rightarrow \overline{\nu}_e$ oscillations might 
also be sensitive to CP--violation in the lepton sector.  Hence, in 
principle the $\nu_\mu \rightarrow \nu_e$ mode at a superbeam can 
offer a handle on much of the physics that the $\nu_e \rightarrow \nu_\mu$
mode offers at a neutrino factory. Indeed, for low energy neutrino beams 
($E_\nu < 10$~GeV) the neutrino fluxes at a superbeam are comparable 
or larger than the corresponding neutrino factory fluxes
(see Table~1). However, there 
is a substantial qualitative difference between searching for $\nu_e$ 
appearance at a superbeam and $\nu_\mu$ appearance at a neutrino factory. 
The signal signature at a superbeam is the appearance of an isolated 
electron (or positron) in a charged current (CC) event. As we will see, 
this signature is plagued with backgrounds at the level of O(1\%) 
of the 
total CC rate. For comparison, the signal signature at a neutrino factory is 
the appearance in a CC event of a wrong--sign muon (a muon of the 
opposite charge--sign to that of the muons stored in the muon ring). 
This signature enables backgrounds to be suppressed to the level of 
O(0.01\%) 
of the total CC rate~\cite{report}. 
Hence, to understand the oscillation 
physics potential at a superbeam we must have a good understanding 
of the backgrounds and the systematic uncertainties associated with 
the background subtraction.

In Section~3 of this report we begin by 
discussing the properties of conventional neutrino beams. 
The motivation for large mass detectors with excellent 
background rejection is discussed in Section~4. 
The most important backgrounds 
to $\nu_\mu \to \nu_e$ and $\bar\nu_\mu \to \bar\nu_e$  
oscillations are discussed in Section~5. 
The physics capabilities of multi--GeV long baseline experiments 
and 1~GeV medium baseline experiments are discussed 
respectively in Sections~6 and 7. 
A summary is given in Section~8. Throughout we will use 
the three--flavor oscillation framework which is reviewed in Appendix~1, 
with the ``leading'' oscillation parameters $\sin^2 2\theta_{23}$ and 
$\delta m^2_{32}$ determined by the atmospheric neutrino deficit, 
and the sub--leading parameters $\sin^2 2\theta_{12}$ and $\delta m^2_{21}$ 
determined by the solar neutrino deficit. 
This choice is appropriate if the LSND effect~\cite{lsnd} is not confirmed (by 
MiniBooNE~\cite{miniboone}, for example). 
Should the LSND effect be confirmed, there is 
likely to be a strong physics case for a low intensity neutrino factory, 
which might be constructed on a relatively fast timescale with only 
a short R\&D phase. The case for a separate superbeam program is less 
obvious (or at least different) in this case.

\begin{table} 
\caption{Neutrino event rates assuming no oscillations, 
compared with intrinsic beam 
backgrounds for conventional and muon-derived beams of 
comparable energies. 
The calculations assume a 1.6~MW proton source is used for the 
MINOS-type beam, the neutrino factories provide 
$2\times 10^{20}$ muon decays per year in the beam--forming straight section, 
and the detector is 732~km downstream of the neutrino source.} 
\vspace{.2in} 
\centerline{
\begin{tabular}{llll}
\hline
Beam
& $<E_\nu>$  & $\nu_\mu$ CC Events & $\nu_e/\nu_\mu$ \\ 
(Signal: $\nu_\mu \to \nu_e$) 
 & (GeV) & (per kton-year) & Fraction \\
\hline 
MINOS-LE &   3.5 & 1800   & 0.012 \\ 
MINOS-ME &   7   & 5760   & 0.009 \\
MINOS-HE &   15  & 12800  & 0.006 \\
\hline 
&&\\
\hline
Beam  & $<E_\nu>$  & $\nu_e$ CC Events & $\nu_\mu/\nu_e$ \\ 
(Signal: $\nu_e \to \nu_\mu$) & (GeV) & (per kton-year) & Fraction \\
\hline 
4.5 GeV $\mu$ Ring  & 3.5 & 400   & 0 \\ 
9.1 GeV $\mu$ Ring  & 7   & 3700  & 0 \\ 
18.2 GeV $\mu$ Ring & 15  & 31400 & 0 \\ 
30 GeV $\mu$ Ring   & 20  & 72600 & 0 \\ 
\hline 
\end{tabular} 
}
\vskip 0.5in
\label{tab:realfoc} 
\end{table}

\section{Beam characteristics and event rates} 

     A conventional neutrino beam is produced using a primary proton 
beam to create a secondary beam of charged pions and kaons, which 
are then allowed to decay to produce a tertiary neutrino beam. 
The secondary beam can be charge--sign selected to produce 
either a neutrino beam from positive meson decays or an antineutrino 
beam from negative meson decays. The 
secondary particles, which are confined radially using either a 
quadrupole channel or horn focusing, are allowed to decay in 
a long decay channel. The resulting neutrino beam consists mostly of 
muon neutrinos (or antineutrinos) 
from $\pi^\pm \rightarrow \mu\nu_\mu$ decays, 
with a small ``contamination'' of electron neutrinos, 
electron antineutrinos, and muon antineutrinos from muon, kaon, and charmed 
meson decays. 
The fractions of $\nu_e$, $\overline{\nu}_e$ and $\overline{\nu}_\mu$ 
in the beam depend critically on the beamline design.  

\begin{figure}
\epsfxsize=\textwidth
%\epsfbox{pikspect.ps}
\epsfbox{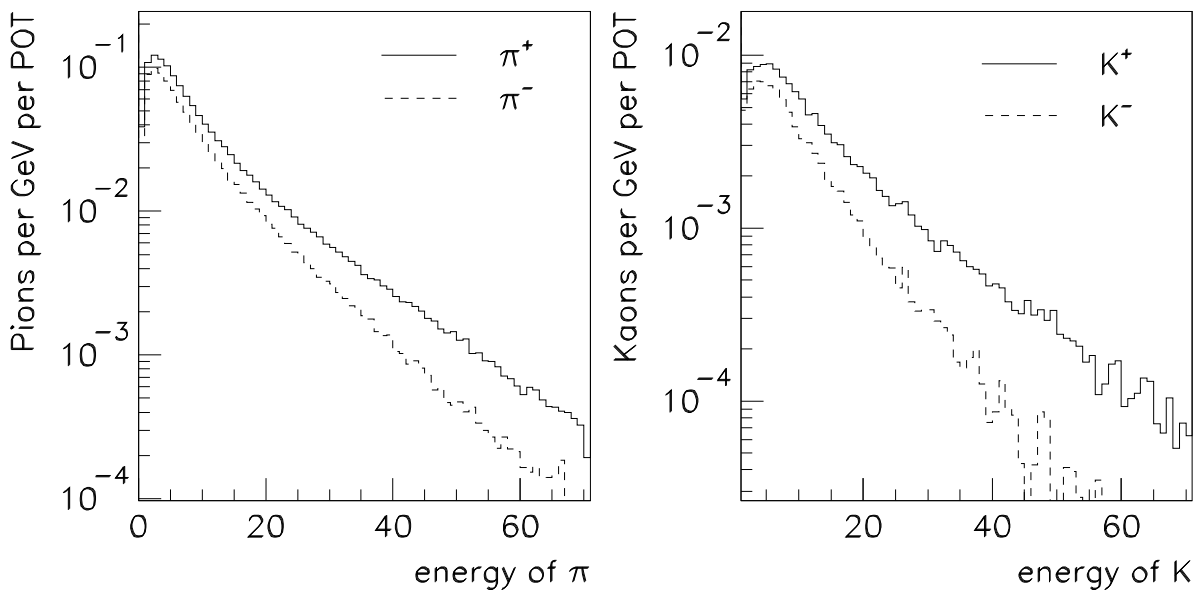}
\caption{Differential spectra for pions (left) and kaons (right) 
produced when 120~GeV protons are incident on the MINOS target. 
The distributions are normalized to correspond to the number of 
particles per incident proton.}
\label{fig:pikspect}
\end{figure}

Figure \ref{fig:pikspect} 
shows the momentum spectrum of charged pions and kaons 
produced when a 120~GeV beam of protons 
strikes a 2~interaction length graphite target~\cite{fpp}.  
Note that at low momenta, positive and negative secondaries are created
at comparable rates, but at higher momentum there is a marked asymmetry. 
For example, at 20~GeV 
the ratio of positive to negative secondaries is 3/2. 
Hence, for high energy beams there is a flux penalty in producing 
an antineutrino beam rather than a neutrino beam. This flux penalty 
increases with increasing beam energy.

\begin{figure}
\epsfxsize=.7\textwidth
%\centerline{\epsfbox{comp_2.eps} }
\centerline{\epsfbox{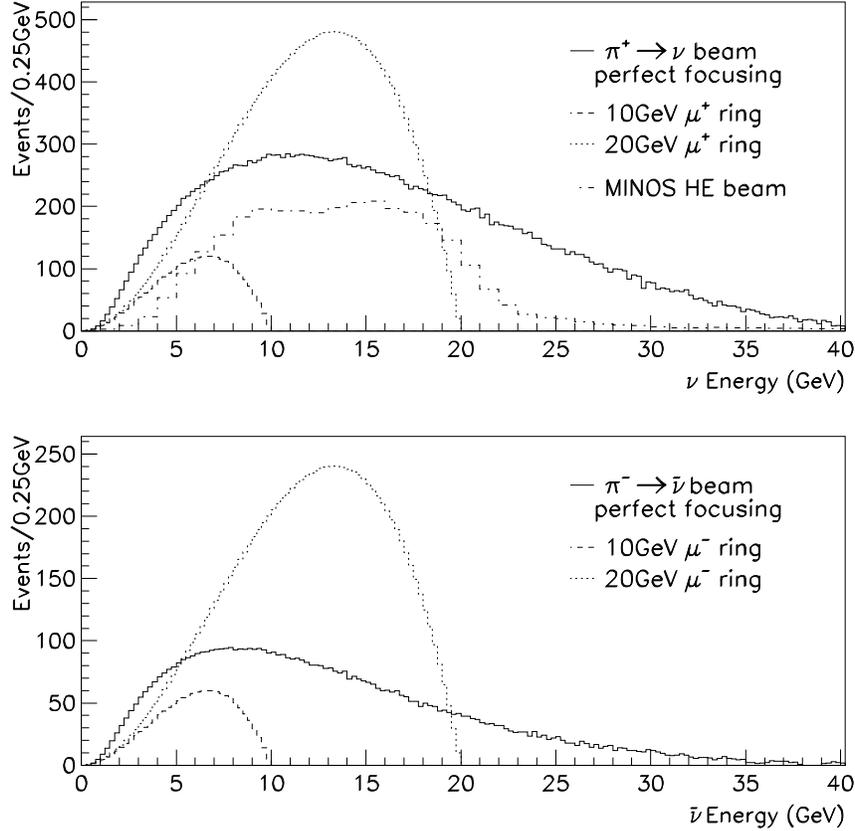} }
\caption{ Differential 
$\nu_\mu$(top) and $\bar\nu_\mu$ (bottom) event rates in the MINOS detector 
732~km downstream of a perfectly focused beam of pions and kaons 
produced when $15 \times 10^{20}$ 120~GeV protons are incident on a graphite 
target ($4 \times$~NuMI for 1~year). 
The top panel also shows the predicted spectrum for a realistic focussing 
system, namely the MINOS HE beam (as indicated). Note that the 
rates are expected to be a factor of 2 or 3 less for a realistic system 
than for a perfectly focussed beam. 
The distributions are compared
to the corresponding $\nu_e$ event rates 732~km downstream of 
10~GeV and 20~GeV muon storage 
rings driven by a 1.6~MW 16~GeV proton source.} 
\label{fig:numu}
\end{figure}

For the two-body decays $\pi\to\mu\nu_\mu$ and $K\to\mu\nu_\mu$ 
there is a one to one correspondence between the energy of the parent
meson and the energy of the neutrino at the far detector.  
For a parent particle of mass $m_h$ and energy 
$E_h$, traveling in a direction $\theta_{\nu h}$ with respect 
to the far detector, the neutrino energy is given by:
\begin{equation}
E_\nu \; = \; 
\frac{m_h^2-m_\mu^2}{2m_h}\times \frac{m_h}{E_h-p_h \cos\theta_{\nu h}} \; 
\approx \; E_h \frac{m_h^2-m_\mu^2}{m_h^2}\frac{1}{1+\gamma^2\theta_{\nu h}^2}
\; .
\end{equation}
For a perfectly focused beam $\theta_{\nu h}=0$, and   
$E_\nu = 0.42 E_h$ for pion decays and $0.95 E_h$ for kaon decays.  
In practice the beamline is designed to focus pions within a 
given momentum window. A broader pion momentum acceptance will 
result in a higher flux of neutrinos 
in the forward direction, but will also result in a  
broader energy spread within the neutrino beam.
The flux of neutrinos per meson at the far detector is given by:
\begin{equation}
\phi \; = \; 
BR \frac{1}{4\pi L^2}\left( \frac{m_h}{E_h-p_h \cos\theta_{\nu h}}
\right)^2 \; \approx \; BR \frac{1}{4\pi L^2}\left( \frac{2\gamma}
{1+\gamma^2\theta_{\nu h}^2} \right)^2 \; ,
\end{equation}
where $BR$ is the branching fraction for the appropriate meson decay, 
$\gamma$ is the Lorentz boost of the decaying particle, 
and $L$ is the distance to the detector.  Note that the 
flux at the far detector has the familiar $\gamma^2$ dependence, 
%that is seen for the case of neutrino beams from a muon storage ring, 
and since the cross--section increases linearly with $\gamma$, 
the event rate has a $\gamma^3$ dependence. 
%be it the pion, kaon, or muon.  
%The big difference between the muon storage 
%ring and conventional beams in terms of raw flux is that where for a 
%muon storage ring one collects as many pions as possible and then focuses 
%the resulting 
%muons to create a monochromatic muon beam, in a conventional beam 
%one designs a beamline to focus the secondary particles in some 
%momentum window, and then lets them decay.  The width of the resulting
%neutrino beam is then just a function of the width of the accepted 
%secondary particles, whereas the width of a neutrino factory beam 
%has only to do with the nature of the three-body muon decay.  
%
Figure~\ref{fig:numu} shows, for a perfectly focussed beam, the 
calculated 
$\nu_\mu$ and $\bar\nu_\mu$ event rates per kton-year for the MINOS 
detector at $L = 732$~km, 
assuming a total decay region of 725~m. 
Note that for a realistic focusing system the neutrino flux would 
be reduced, typically by a factor of 2 or 3. 
The calculation shown in Fig.~\ref{fig:numu} is for  
$15\times 10^{20}$ 120~GeV protons striking a graphite target
(4 x NuMI~\cite{numitdr} for 1~year) 
similar to the one being constructed for the MINOS 
beam~\cite{fpp}, with the beamline set to focus either positive or 
negative pions. 
Also shown are the corresponding event rates at a neutrino 
factory in which there are $2 \times 10^{20}$ useful muon decays, 
with $E_\mu = 10$ and 20~GeV. 

The total CC event rates for 1.6~MW NuMI low, medium, and high energy 
superbeams are compared in 
Table~\ref{tab:realfoc} with the corresponding (same $<E_\nu>$) 
rates at a neutrino factory. 
Note that the rates at a neutrino factory rapidly exceed the 
corresponding conventional beam rates for neutrino beam energies 
exceeding about 10~GeV. For lower energies, conventional beams 
provide higher rates although the 
beam is not background free, and the electron appearance signal 
is experimentally challenging.
%
%\begin{table} 
%\label{tab:realfoc} 
%\caption{Expected neutrino CC event rates (per kt-year),
% in the absence of oscillations, at $L = 732$~km. 
%Rates at 1.6~MW NuMI superbeams (PERFECT FOCUSING ?) 
%are compared with the corresponding 
%rates at a neutrino factory delivering $2\times 10^{20}$ useful muon 
%decays.  THIS IS THE NORMALIZATION WE WANT.
%} 
%\begin{tabular}{lllll} 
% & $<E_\nu>$  & \multicolumn{2}{c}{Events/kt-yr} & $E_\mu$ \\ 
%Beam & (GeV) & $\pi$  & $\mu$ & (GeV) \\
%\hline\hline 
%MINOS-LE &   3.5   & 1800   & 120   & 4.5 \\ 
%MINOS-ME &   7   & 5760   & 1120  & 9.1 \\
%MINOS-HE &   15 & 12800  & 9430 & 18.2 \\
%         &   20 &        & 21770 & 30 \\
%\end{tabular} 
%\end{table} 

\begin{figure}
\hspace{-.5in} 
\epsfxsize=.5\textwidth 
%\centerline{\epsffile[10 30 510 540]{eff_vs_l.ps} }
\centerline{\epsffile[10 30 510 540]{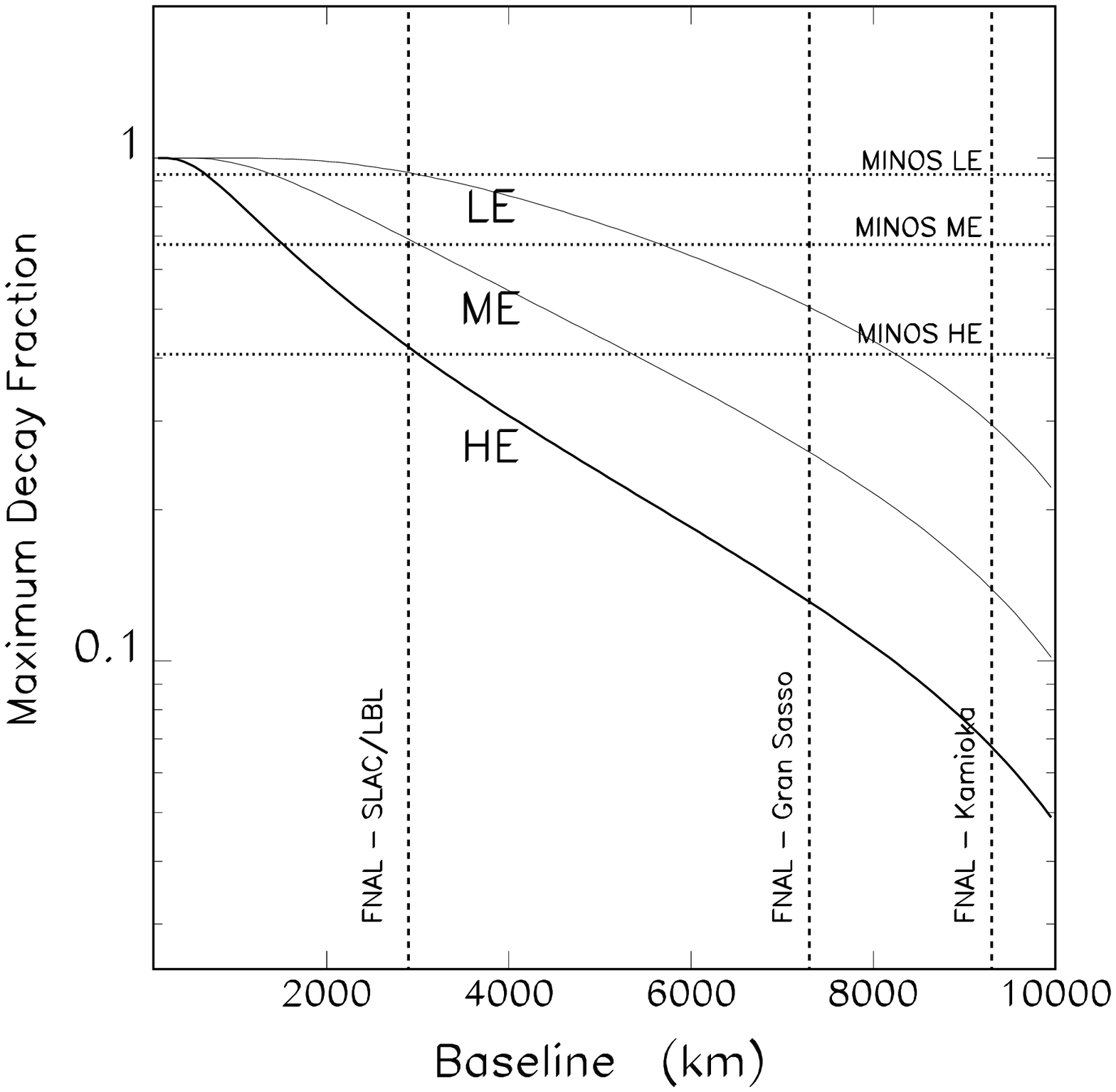} }
\caption{Fraction of $\pi^\pm$ that decay in a channel of maximum 
length within a 200~m deep rock layer, shown as a function of 
baseline for three different average NuMI beam energies: 
3.5~GeV (LE), 7~GeV (ME), and 15~GeV (HE). 
For convenience, distances from Fermilab to SLAC/LBNL, Gran~Sasso, and Kamioka 
are indicated by vertical broken lines, and the decay fractions 
for the foreseen NuMI beams ($L = 730$~km) are indicated by dotted 
horizontal lines.} 
\label{fig:flxvbsln}
\end{figure}

Finally, the decay channel for a very long--baseline superbeam 
must fit within the viable rock layer below the proton driver. 
At Fermilab this rock layer is $\sim 200$~m deep, below which 
there is a deep aquifer. This limits the  
length of the decay pipe to less than 200m$\times \sin \theta$, 
where the dip angle $\theta$ depends on the baseline: 
$L = 12756 \times \sin\theta$~km.
In fact the length of the decay region is further restricted by 
the depth of rock used to bury the proton driver, bend the proton 
beam to the required direction, accommodate the target and 
focusing systems, and if there is a near detector, accommodate the 
associated shielding and near detector hall. 
Figure~\ref{fig:flxvbsln} shows the relative flux 
loss for different energy beamlines as a function of baseline 
length~\cite{geersb}. 
The calculation allows 30~m for the near detector shielding plus hall. 
For a baseline of 732~km, 
$\theta = 3.3^\circ$ and the NuMI beam pipe length of 
675~m is not restricted by the depth of the good rock. 
However for a far site at 
7300~km (Fermilab $\rightarrow$ Gran Sasso), $\theta = 35^\circ$, 
and the length of the decay channel is severely restricted, 
so that for the LE (ME) [HE] NuMI beams only 50\% (25\%) [13\%] 
of the pions would decay in the channel.
Clearly, decay channel length 
restrictions must be taken into account in calculating the fluxes 
for very long baselines ($> 3000$~km). 
For trans--Atlantic or trans--Pacific baselines 
there is a premium on minimizing the shielding and detector hall 
length for the near detector, minimizing the length of the targeting 
hall, using high--field dipoles to rapidly bend the 120~GeV proton beam 
to the required direction, and perhaps considering a ``roller-coaster'' 
geometry for the proton beam.

%\subsection{Beamline Cost Considerations}
%
%     One of the practical considerations for an upgraded neutrino 
%beam at Fermilab is related to the construction of a new beam line.  
%The NuMI beam is located in a large layer of dolomite, which is 
%20m - 200m below the surface.  
%The cost of shielding to protect water flowing in 
%the dolomite is a substantial fraction of the \$80M cost of the 
%beam.  In addition, the NuMI beam may be near the Fermilab annual limit 
%for radiation in the ground-water, so a more intense beam would 
%require additional shielding, particularly near the beam pipe and 
%around the target pile.
%     Below 200m, there is a large aquifer,\cite{lach} and the cost of 
%shielding the beam in this alternate layer would be different by a 
%large unknown factor.  There are other engineering challenges for 
%building a beam at a large slant.  One estimate for the relative 
%cost of a steep beam can be found in Reference \cite{1991cdr} 
%where the incremental cost of a beam at 6239 km is 1.8 times 
%larger than a similar beam aimed at 732 km.  
%     Once the NUMI beamline turns on the shielding costs of an 
%upgraded conventional beam will be better known.  The NUMI 
%beamline shielding needs have themselves been unprecedented, 
%and as such the amount of shielding NUMI uses has additional 
%margins of error beyond the standard safety factors.  

\section{Large detectors and low backgrounds}

Our first physics goal is the observation of $\nu_\mu \rightarrow \nu_e$ 
oscillations. Since, to a good approximation the oscillation probability is  
proportional to the amplitude parameter $\sin^2 2\theta_{13}$, it is useful 
to define the $\sin^2 2\theta_{13}$-reach for a given experiment, which we 
define as that value of $\sin^2 2\theta_{13}$ which would result in a signal 
that is 3 standard deviations above the background. 
Taking the atmospheric neutrino deficit oscillation scale $\delta m^2_{32}$ 
to be in the center of the region indicated by the SuperKamiokande 
(SuperK) data, 
for a given proton driver, superbeam design, and baseline, 
we can calculate the $\sin^2 2\theta_{13}$ reach once we specify 
(a) the data sample size $D$ (kt--years), defined as the product of 
the detector fiducial mass, the efficiency of the signal selection 
requirements, and the number of years of data taking, 
(b) the background fraction $f_B$, defined as the background rate divided 
by the total CC rate for events that pass the 
signal selection requirements, and 
(c) the fractional uncertainty $\sigma_{f_B}/f_B$.
Note that $D$ determines the statistical uncertainty on 
the signal. For a fixed $D$, $f_B$ determines the statistical uncertainty 
on the background, and $f_B \times \sigma_{f_B}/f_B$ determines 
the systematic uncertainty on the background subtraction.

\begin{figure}
%\hspace{.5in}
%\vspace{1.0cm}
\epsfxsize=.85\textwidth 
%\epsffile[0 0 540 685]{jhf_contours.ps}
\epsffile[0 0 540 685]{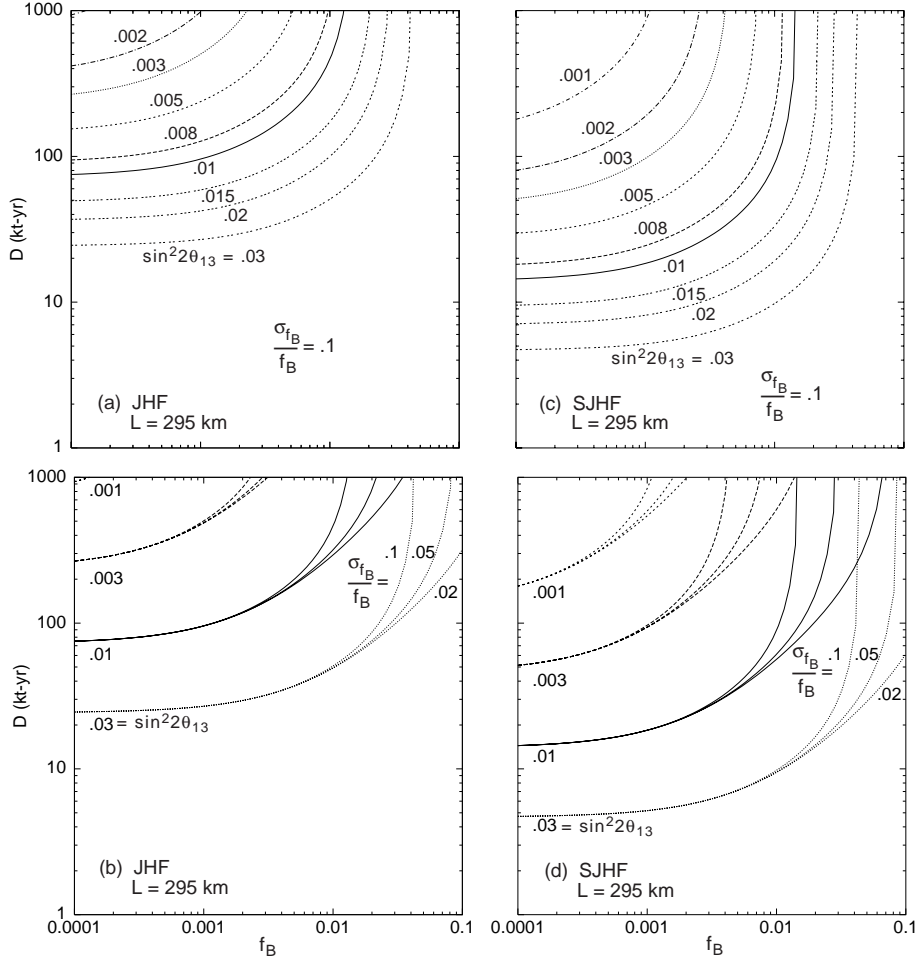}
\vspace{-4.0cm}
\caption{
Contours of constant $\sin^22\theta_{13}$ reach that 
correspond to a $\nu_e\to \nu_\mu$ signal that is 3 standard deviations 
above the background~\cite{geersb}. 
The contours are shown in the $(D, f_B)$--plane, 
where $D$ is the data-sample size and $f_B$ the background rate divided 
by the total CC rate. 
The contours are shown for 
the 0.77~MW (left-hand plots) and 4.0~MW (right-hand plots) 
JHF scenarios with
$L = 295$~km. The top panels show curves for $\sigma_{f_B}/f_{B} =
0.1$, while the bottom panels show curves for $\sigma_{f_B}/f_{B} =
0.1$, $0.05$, and $0.02$.
} 
\label{fig:jhfcontours}
\end{figure}

\begin{figure}
\hspace{-.5in} 
\epsfxsize=.75\textwidth 
%\centerline{\epsffile[10 30 510 540]{snumi_contours.ps} }
\centerline{\epsffile[10 30 510 540]{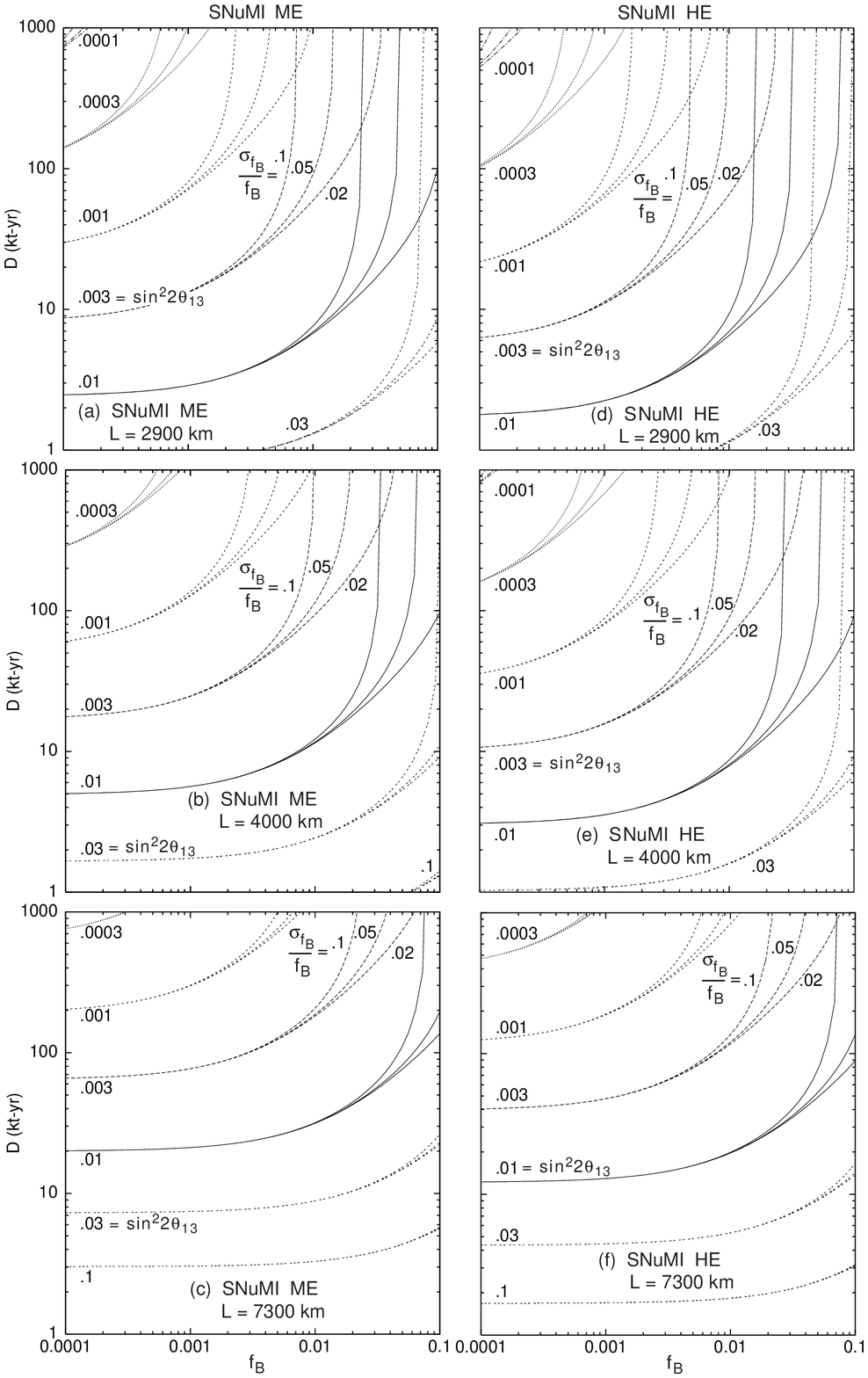} }
\vspace{1.0cm}
\caption{Contours of constant $\sin^22\theta_{13}$ reach~\cite{geersb} that 
correspond to a $\nu_e\to \nu_\mu$ signal that is 3 standard deviations 
above the background, at upgraded 1.6~MW NuMI ME (left) and HE 
(right) beams. 
The contours are shown in the $(D, f_B)$--plane, 
where $D$ is the data-sample size and $f_B$ the background rate divided 
by the total CC rate. 
The contours are shown for 
$L = 2900$ (top), $4000$ (center), and $7300$~km (bottom). 
Curves are shown for
systematic uncertainties on the background subtraction 
$\sigma_{f_B}/f_{B} = 0.1$, $0.05$, and
$0.02$.
} 
\label{fig:contours}
\end{figure}

Contours of constant $\sin^2 2\theta_{13}$ reach 
that correspond to various values of $\sigma_{f_B}/f_B$  
are shown in the 
$(f_B,D)$--plane in Fig.~\ref{fig:jhfcontours} 
for $E_\nu = 1$~GeV superbeams produced using 0.77~MW and 4~MW 
JHF proton drivers. 
The corresponding contours are shown in Fig.~\ref{fig:contours} 
for long--baseline 
1.6~MW medium energy and high energy NuMI superbeams. 
The contours have a characteristic shape. 
At sufficiently large $D$ the $\sin^2 2\theta_{13}$ sensitivity 
is limited by the systematic uncertainty on the background subtraction,
and the reach does not significantly improve with increasing dataset size. 
The contours are therefore vertical in this region of the figures. 
At sufficiently small $D$ the sensitivity of the $\nu_\mu \rightarrow \nu_e$ 
appearance search is limited by signal statistics, and further reductions in 
$f_B$ do not improve the $\sin^2 2\theta_{13}$ reach. The contours are 
therefore horizontal in this region of the $(f_B, D)$--plane. 
Note that the next 
generation of neutrino experiments are expected to achieve 
$\sin^2 2\theta_{13}$ 
reaches of $\sim 0.03$. Now consider as examples the 4~MW JHF beam, 
and the 1.6~MW NuMI beam with $L = 2900$~km. 
Taking $\sigma_{f_B}/f_B = 0.05$, 
an inspection of the figures leads us to conclude 
that if we wish to improve the $\sin^2 2\theta_{13}$ sensitivity an order 
of magnitude beyond the next generation of experiments, 
in the limit of very massive detectors at the JHF (NuMI) superbeams, we 
can only tolerate background fractions $f_B < 0.007$ (0.004). 
Achieving the required background rejection in a water cerenkov detector 
seems challenging. Achieving the required detector masses with  
detector technologies that can meet the $f_B$ requirements also 
seems challenging. Hence an understanding of the physics capabilities at 
a superbeam necessarily begins with an understanding of the 
parameters $D$, $f_B$, and $\sigma_{f_B}/f_B$ that can be achieved 
with realistic (but futuristic) detectors.

In the following section we consider the backgrounds, and for realistic 
futuristic detectors, $D$ and $f_B$. This will enable us to use the 
$(f_B,D)$ figures to determine the $\sin^2 2\theta_{13}$ sensitivity 
at various superbeams.

\section{Backgrounds to $\nu_\mu \rightarrow \nu_e$ oscillations} 

Backgrounds 
play a critical role in determining the sensitivity of a superbeam experiment 
to $\nu_\mu \rightarrow \nu_e$ oscillations.  
In our neutrino factory studies~\cite{report} 
it was straightforward to obtain backgrounds to the $\nu_e\to\nu_\mu$ 
search at the $f_B \sim 10^{-4}$ level by cutting on the final state 
muon momentum. 
However, for conventional neutrino beams 
reducing the background rates below $10^{-2}$ of the total CC rate is not 
trivial and  
would require significant reductions in the beam flux and signal selection 
efficiency. 
There are four important sources of background in a 
$\nu_e$ appearance search at a superbeam:  
(i) electron--neutrinos produced in the initial beam, 
(ii) neutral current (NC) neutrino interactions in which a $\pi^0$ is 
mis--identified as an electron, 
(iii) CC $\nu_\mu$ interactions in which a $\pi^0$ is 
mis--identified as an electron, and the muon is not identified, and 
(iv) events from $\nu_\mu \to \nu_\tau$ oscillations followed by $\nu_\tau$ 
CC interactions and either $\tau \to e+X$ decays, or 
$\tau \to \pi^0 + X$ decays in which the $\pi^0$ fakes an electron. 
In the following we discuss these backgrounds and how, for 
different detector technologies, the backgrounds might be suppressed. 

\subsection{Electron--neutrino contamination} 

In a conventional neutrino beamline, muon neutrinos are produced in  
two-body pion-- and kaon--decays.  However, the charged and neutral kaons can 
also undergo 3--body decays to produce an electron neutrino (or antineutrino). 
Furthermore, secondary muons in the beamline  
can also decay to produce electron neutrinos.  Generally speaking, higher 
primary proton beam energies yield higher kaon/pion ratios,  
and longer decay channels permit more muon decays.
The expected electron neutrino background fractions are listed in 
Table~\ref{tab:nuefrac} for  
the next generation of conventional neutrino beams. Note that for 
most of the beamlines listed the $\nu_e$ background fraction 
is around the $1\%$ level. The exceptions are the MiniBooNE beam (which 
benefits from a relatively low primary proton energy), and 
ORLaND~\cite{orland} (which 
uses stopped muons).  
Since the electron neutrinos within the beam are created in three--body 
decays, the $\nu_e$ energy spectrum is typically much 
broader than the $\nu_\mu$ spectrum.
This effectively means that there is always 
some fraction of the electron neutrino flux which overlaps in energy with the 
muon neutrino flux.
\begin{table}
\caption{Electron neutrino fractions and the fractional energy 
spreads for a selection of 
current (or next) generation conventional neutrino beams.
Note that most beamlines produce a beam with a fractional 
energy spread between 30\% and 50\%, and a $\nu_e$ 
contamination that ranges from 0.2\% to 1.2\%, for beams at or 
above 1~GeV. }  
\centerline{    
\begin{tabular}{lcccc} 
&&&&\\
\hline
Beamline & Proton & Peak $\nu_\mu$  & $\nu_e/\nu_\mu$ &  
$\sigma_{E_\nu}/E_\nu$ \\
 & Energy (GeV) & Energy (GeV) & ratio &  \\ 
\hline 
K2K      & 12 & 1.4 & 0.7\%  &  1.0 \\
MINOS LE & 120 & 3.5 & 1.2\%  &  0.28 \\
MINOS ME & 120 & 7  & 0.9\%  & 0.43 \\
MINOS HE & 120 & 15 & 0.6\%  & 0.47 \\
CNGS     & 400 & 18 & 0.8\%  & 0.33 \\
JHF   wide  & 50  & 1  & 0.7\% & 1.0  \\
JHF   HE    & 50  & 5  & 0.9\% & 0.40 \\
MiniBooNE   & 8   & 0.5& 0.2\% & 0.50  \\
ORLaND      & 1.3 & 0.0528 & 0.05\% & 0.38 \\
\hline
\end{tabular} }
\vskip 0.5in
\label{tab:nuefrac} 
\end{table}      

%\begin{table}
%\label{tab:nuefrac} 
%\caption{Electron neutrino fractions for a selection of 
%the current (or next) generation of conventional neutrino beamlines.}      
%\begin{tabular}{lllllll} 
%Beamline & Peak $\nu_\mu$  & $\nu_e/\nu_\mu$ & p Energy \\
%        & Energy (GeV) & event ratio & GeV \\ 
%\hline 
%K2K      & 1.4 & 0.7\%  & 12 \\
%MINOS LE & 3.5 & 1.2\%  &  120 \\
%MINOS ME & 7  & 0.9\%  & 120 \\
%MINOS HE & 15 & 0.6\%  & 120 \\
%CNGS     & 17 & 0.8\%  & 400 \\
%JHF   wide  & 1 & 0.7\% & 50 \\
%JHF   HE  & 5  & 0.9\% & 50 \\
%MiniBooNe & 0.5  & 0.2\% & 8 \\
%ORLaND & 0.0528 & 0.05?\% & 1.3 \\
%\end{tabular} 
%\end{table}      

\begin{figure}
\centerline{
  \epsfxsize=.45\textwidth
%  \epsfbox[0 0 369 369]{nuebk.eps}
  \epsfbox[0 0 369 369]{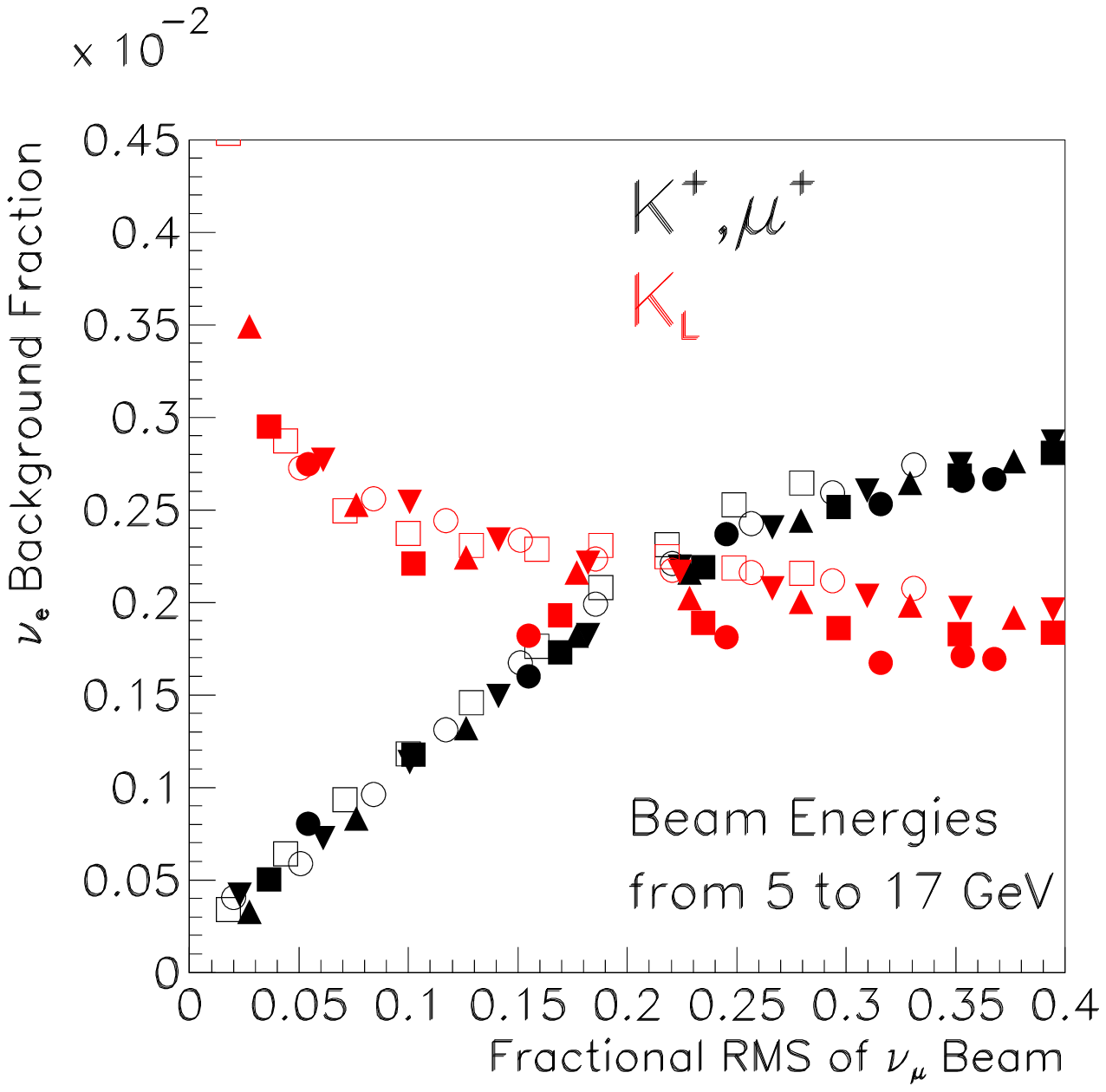}
  \epsfxsize=.45\textwidth
%  \epsfbox[0 0 369 369]{nubarebk.eps}
  \epsfbox[0 0 369 369]{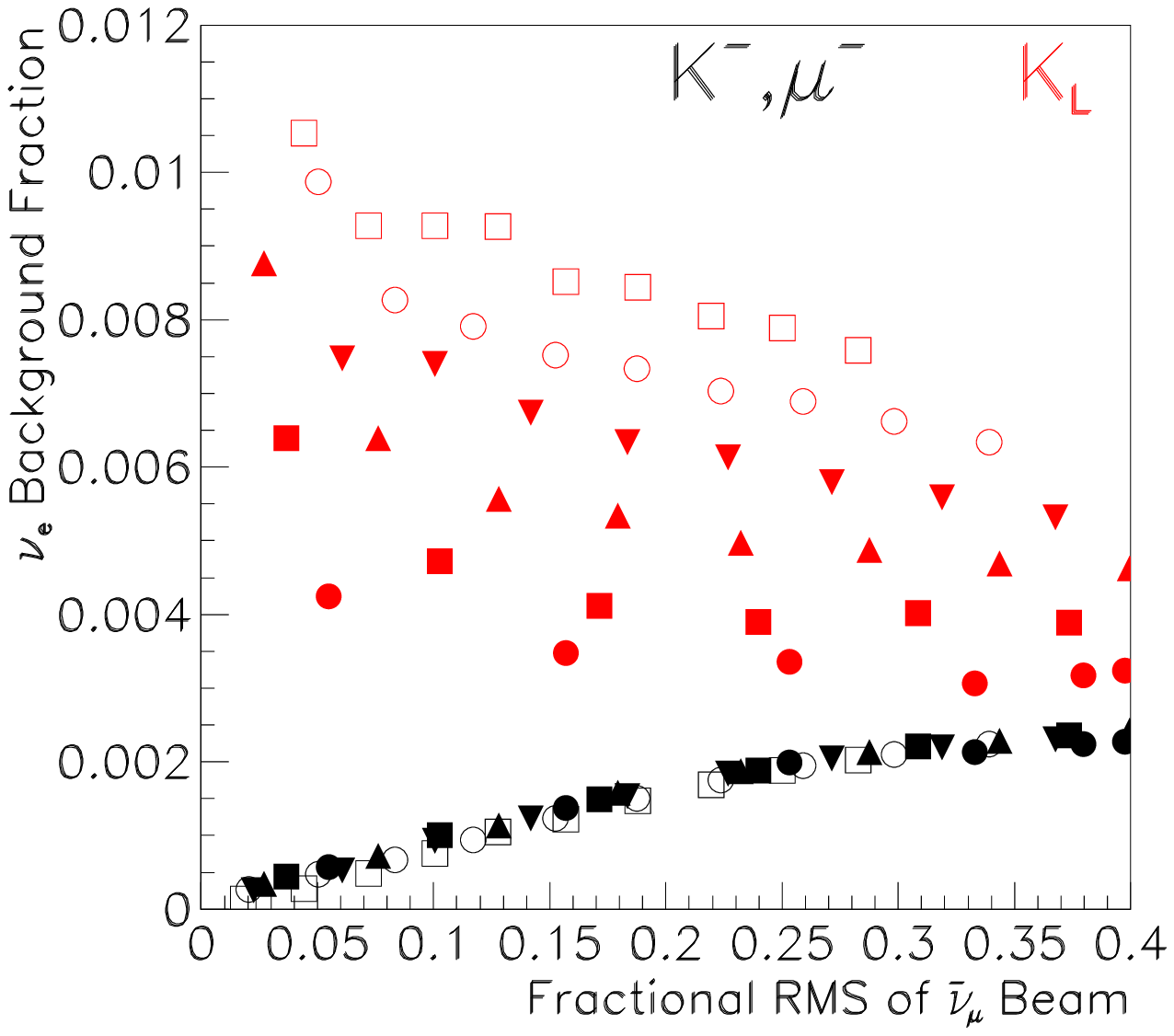} 
}
\caption{Fraction of $\nu_e+\bar\nu_e$ events in a $\nu_\mu$ (left) 
or $\bar\nu_\mu$ (right) beam shown 
as a function of the fractional beam momentum spread.  The various symbols 
correspond to different mean beam energies, from 5~GeV (filled circles) 
to 17~GeV (open squares).
The distributions that increase (decrease) from left to right are the 
contributions from $K^\pm, \mu^\pm$ ($K_L$) decays.
}  
\label{fig:nuebkgd} 
\end{figure}

The $\nu_e$ flux contributions from $K^\pm$ and $\mu^\pm$ decays 
can be reduced by decreasing the secondary particle momentum acceptance. 
The contributions from $K_L$ decays can be reduced by 
putting large bends in the beamline immediately after the proton target. 
The neutral meson backgrounds 
are particularly dangerous for $\bar\nu$ running because they give 
electron neutrinos in a $\bar\nu_e$ appearance search, 
and the cross section for the $\nu_e$ background is therefore twice 
as large as for the $\bar\nu_e$ signal. 
It has been suggested~\cite{richter} that a neutrino beam could be made 
with an extremely small $\nu_e$ contamination by only 
accepting pions and kaons within a narrow momentum interval, 
and then rejecting 
neutrino events that have a total energy inconsistent with the 
expected neutrino beam energy. To understand how this might work we 
consider the correspondence between the  
the neutrino beam energy spread and the contribution to $f_B$ 
from the initial $\nu_e$ flux, which comes from  $K^\pm$, $\mu^\pm$, 
and $K_L$ decays. 
The ratio of the initial $\nu_e$ flux to $\nu_\mu$ flux is shown 
as a function of the fractional beam energy spread in 
Fig.~\ref{fig:nuebkgd} for a perfectly focussed secondary beam 
produced with 120~GeV primary protons 
on a 2~interaction length graphite target, followed by 
a 725~m long decay channel~\cite{fpp}.  
To achieve a background fraction that is no larger than 0.1\% 
in either $\nu$ or $\bar\nu$ running would require a beamline 
with a momentum acceptance no larger than 10\%. In addition,  
we must suppress the remaining $\nu_e$ contribution from $K_L$ 
decays (using one or more dipoles after the proton target) 
by factors in excess of 2.5 and 5 for $\nu$ and $\bar\nu$ 
running respectively. 
Noting that the effective $\nu_\mu$ flux also decreases roughly 
linearly with the momentum acceptance, we conclude that, 
even for an idealized perfectly focussed beam, reducing 
the initial $\nu_e$ contamination in the beam to 0.1\% will result 
in an order of magnitude reduction in $D$.

An alternative way of decreasing the $\nu_e$ contamination from kaon decays 
is to decrease the primary proton energy (from 120~GeV to something less). 
However, the MiniBooNE beam, which uses 8~GeV primary protons, 
is expected to achieve a $\nu_e$ contamination of no better than 0.2\%.
Furthermore, even if kaon contributions are completely eliminated, muon decays 
still provide electron neutrinos.
%
%However, the number of high energy pions in the secondary beam that are 
%selected within a fixed momentum 
%interval also decreases with decreasing proton beam energy.
%Overall the number of kaons per pion falls 
%like the proton energy to the XXX power, for a perfectly focused
%beam (***possible plot here comparing k/pi ratio vs proton energy***). 
%NOT SURE THIS IS TRUE - DONT FORGET LOWER ENERGY MEANS FASTER MACHINE CYCLE.

%The tradeoff between event rate and electron neutrino contamination 
%depends on many variables. 
%A lower primary proton momentum would 
%reduce the contamination from kaon decays, but the lower one goes in 
%proton momentum the lower the fraction of pions produced to give one that 
%neutrino energy.  The numbers shown in the plots above are for 
%a beam made with 120GeV protons, going to lower initial proton momentum 
%would reduce the kaon contribution significantly.  However, to focus the 
%same energy pions for a lower momentum proton beam gives a much reduced pion 
%flux, since there is approximate scaling between the pion energy and 
%the proton energy.  
%Overall the number of kaons per pion falls 
%like the proton energy to the XXX power, for a perfectly focused
%beam (***possible plot here comparing k/pi ratio vs proton energy***).  

In summary, for a multi--GeV neutrino beam, 
we would not expect a reduction in the momentum acceptance 
of the decay channel to significantly reduce the 
$\nu_e$ contamination in the beam unless 
we are willing to accept a large reduction in $D$. 
%The $\nu_e$ contamination in the NuMI beamline could be 
%reduced without reducing $D$ significantly by using one or more 
%dipoles after the proton target to 
%remove most of the $K_L$ contribution. This would 
%reduce the $\nu_e$ fraction by roughly a factor of three to four.
Optimistically, the contribution to $f_B$ from the $\nu_e$ contamination 
in the beam might be reduced to 
$\sim0.2 - 0.5\%$ for the multi--GeV neutrino beams considered
later in this report.

\subsection{Neutral Current Backgrounds} 

For high energy neutrinos the NC cross section is
roughly 40\% 
of the CC cross section, and is independent
of the initial flavor of the interacting neutrino.  A NC interaction 
can result in a total visible energy anywhere from zero to
the initial neutrino energy $E_\nu$. Some of the visible energy may be 
in the form of neutral pions, which can fake a single electron signature. 
The probability for a NC event to produce an energetic $\pi^0$ that 
fakes an electron depends both on the $\pi^0$ production rate and on 
the details of the detector and signal selection requirements.  
In the following we will begin by considering $\pi^0$ production in 
NC events, and then consider various detector strategies for reducing 
the background.

\subsubsection{Energetic $\pi^0$ production}

Consider the probability that a NC interaction will 
produce an event with an energetic $\pi^0$ that could fake a 
$\nu_e$ CC interaction. 
Our calculations have been performed using the GEANT Monte Carlo 
program together with a LEPTO 
simulation to generate neutrino interactions~\cite{fmcgeant}. 
A rough approximation has been used for quasi-elastic and resonance 
production in CC events, but no resonance production has been 
included for the NC events. 
Note that the lower the neutrino energy, the higher the fraction of 
quasi-elastic and resonance interactions.  
As a benchmark, for a 5~GeV neutrino beam the 
quasi-elastic and resonance contributions are about 33\% of the
total event rate. 
%Since the neutral 
%pions tend to be produced promptly, the rates are not expected to 
%depend much on the neutrino target material.

\begin{figure}
\epsfxsize=.5\textwidth
%\centerline{\epsffile[0 0 515 515]{new_em_dist.eps}}
\centerline{\epsffile[0 0 515 515]{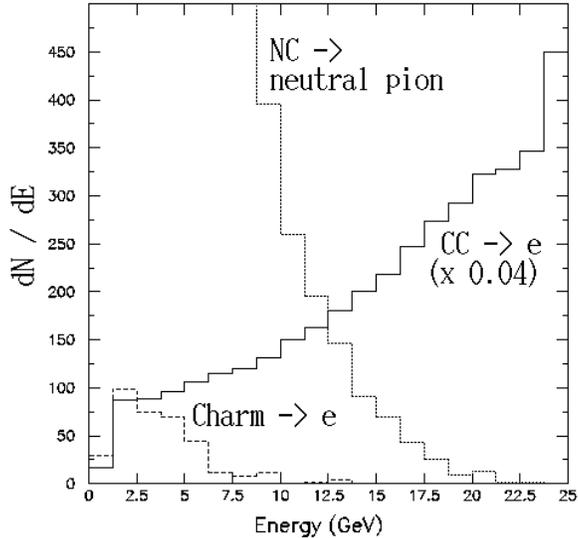}}
\caption{Energy distributions for electrons and neutral pions produced in 
25 GeV neutrino interactions.  Solid line:  
Electrons produced in CC interactions (scaled by 0.04).
Dotted line:  Neutral pions
generated in NC interactions.  Dashed line:  Electrons from charm
semileptonic decay.}
\label{em_dist}
\end{figure}

\begin{figure}
\epsfxsize=.5\textwidth
%\centerline{\epsffile[43 43 457 457]{new_pi0frag.eps}}
\centerline{\epsffile[43 43 457 457]{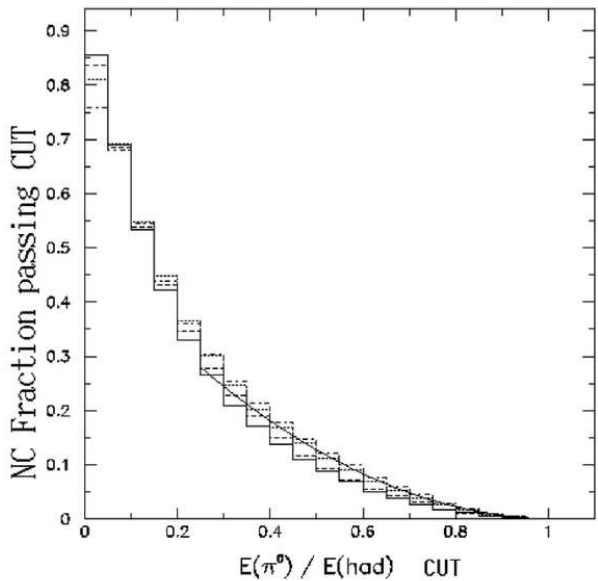}}
\caption{Fraction of NC events with a $\pi^0$
with energy greater than a given fraction of the hadronic energy.
The different histograms are for different ranges of hadronic energy:
Solid:  $20<E_{\rm had}<25$~GeV.  Dashed:  $15<E_{had}<20$~GeV.  
Dotted:  $10<E_{\rm had}< 15$~GeV.
Dot-Dashed:  $5<E_{\rm had}< 10$~GeV. 
The curve shows the parameterization given in the text.
}
\label{pi0frag}
\end{figure}

The simulated energy distribution for electrons produced in 25~GeV 
$\nu_e$ CC events   
is compared in Fig.~\ref{em_dist} with the corresponding 
distribution for $\pi^0$'s produced in 25~GeV NC events.   
The $\pi^0$ rates are small at high energy. A cut on the 
energy of the electron candidate can therefore reduce the 
NC background in a $\nu_\mu \to \nu_e$ search. 
The most dangerous background events 
are those in which the $\pi^0$ takes most of the energy of 
the hadronic fragments, and hence $x \equiv E_{\pi^0}/E_{\rm had}$ is large. 
These neutral pions can fake an isolated electron from a 
$\nu_e$ CC interaction. The fraction of NC events having a $\pi^0$ 
with energy greater than a given fraction of $E_{\rm had}$ 
is shown in  Fig.~\ref{pi0frag} for different ranges of hadronic energy.  
The $\pi^0$ fragmentation functions are roughly independent of $E_{\rm had}$. 
To a good approximation, at large $x$ ($\gsim 0.3$) 
a single function describes all the curves shown in the figure: 
\begin{equation}
p(x) = (0.49)-(0.96)x+(0.47)x^2
\label{pi0prob}
\end{equation}
Note that in a NC event $E_{\rm had}$ is just the visible energy. 
Within the framework of the parton model,  
for a given neutrino energy, the 
visible energy spectrum is described by~\cite{bargergeer}:
\begin{equation}
\frac{1}{N}\frac{dN}{dy} = \frac{15}{16}\left(1+\frac{(1-y)^2}{5}\right) \; .
\label{dsdy}
\end{equation}
where $y = E_{\rm had} / E_\nu$.


\begin{figure}
\centerline{
\epsfxsize=.45\textwidth 
%\epsffile[50 60 515 515]{sb_fritz.eps}
\epsffile[50 60 515 515]{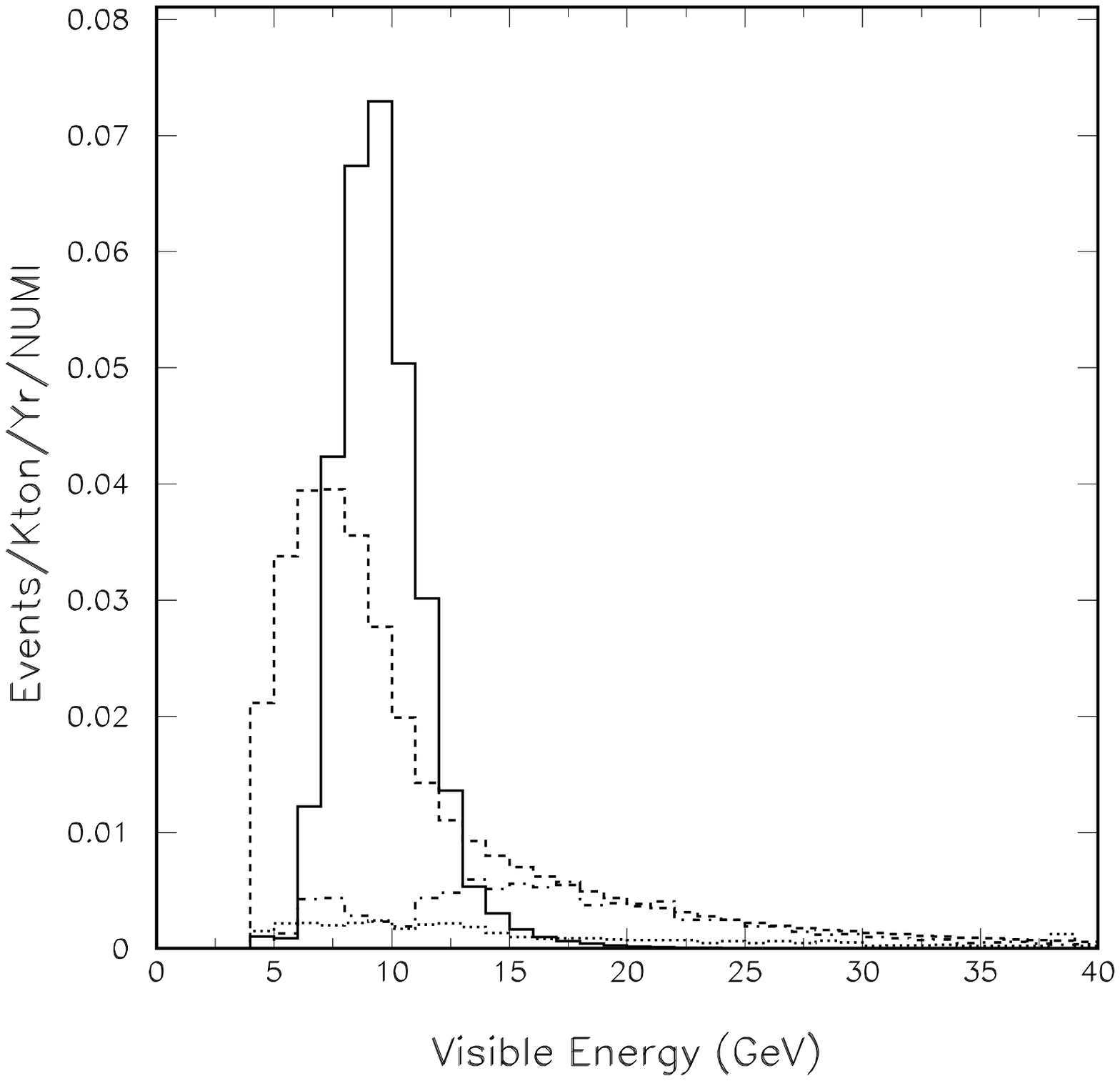}
\epsfxsize=.45\textwidth 
%\epsffile[50 60 515 515]{sb_8_10.eps}
\epsffile[50 60 515 515]{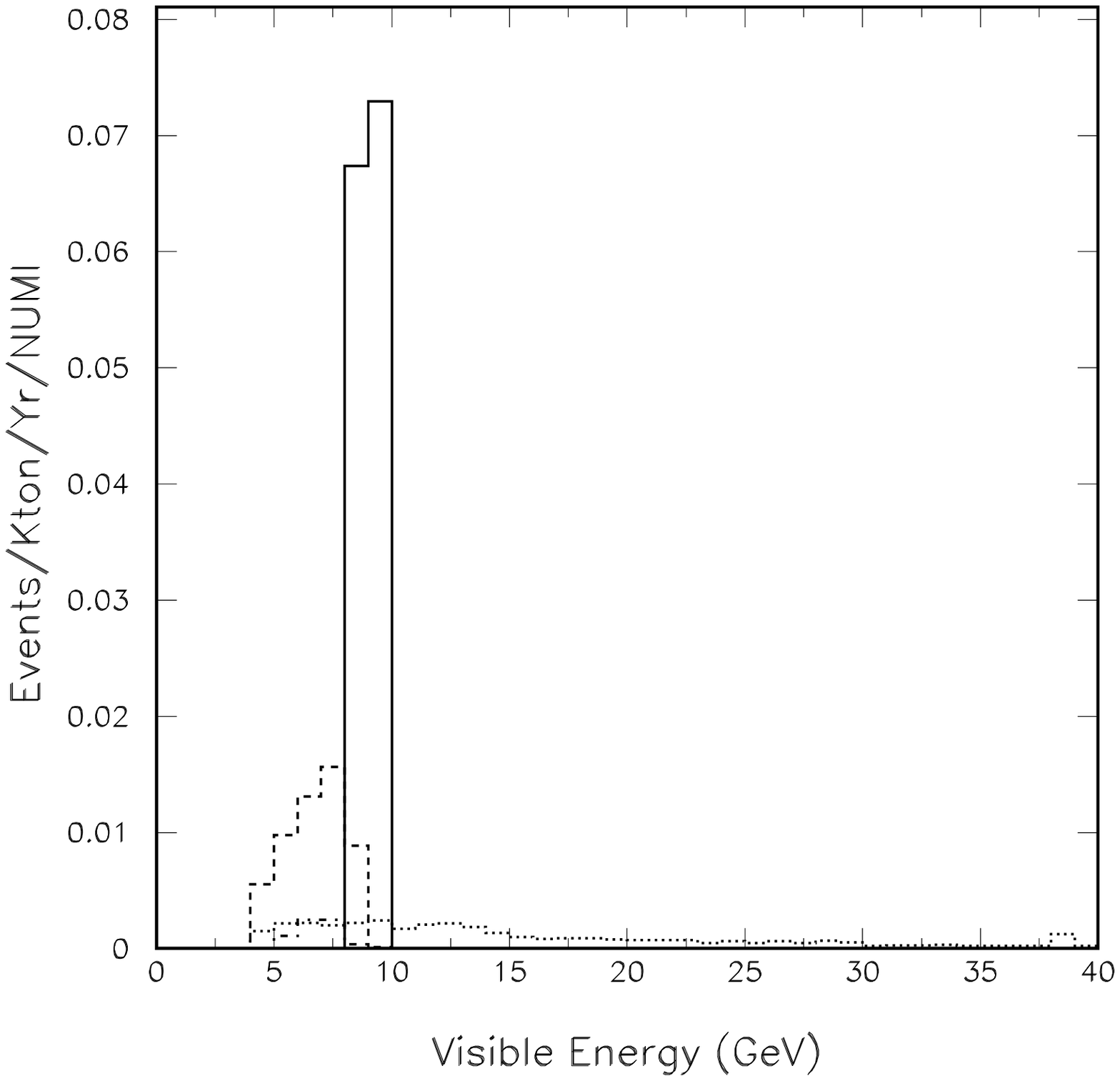}
}
\caption{Differential distributions for NC, $\nu_\tau$, and intrinsic 
$\nu_e$ backgrounds 
after requiring the electron candidate has at least 30\% of the 
total event energy, compared 
with corresponding distributions for a $\nu_\mu \to \nu_e$ signal 
($\sin^2 2\theta_{13} = 0.01$, $\delta m^2_{32} = 3.5 \times 10^{-3}$~eV$^2$) 
at $L = 9300$~km. In the left panel a NuMI-like beam with a mean energy of 
10~GeV is assumed. The right panel shows only the contributions 
from neutrinos in the beam with energies between 8~GeV and 10~GeV. 
Solid line:  CC $\nu_e$ signal. 
Dashed line:  NC background. 
Dot-Dashed line:  $\nu_\tau$ background. 
Dotted line:  Intrinsic $\nu_e$ background.  }
\label{sb}
\end{figure}

We can use Eqs.~\ref{pi0prob} and \ref{dsdy} 
to calculate the probability that there
will be a $\pi^0$ above some cut-off energy.  Thus, for a given 
input neutrino energy spectrum, an output NC background 
spectrum can be calculated under the assumption that all 
of the $\pi^0$'s are incorrectly identified as electrons. 
As an explicit example, consider a two--horn (NuMI--like) neutrino 
beam with a mean energy of 10~GeV, and a baseline of 9300~km 
(Fermilab $\rightarrow$ SuperK), and assume the oscillation 
amplitude parameter $\sin^2 2\theta_{13} = 0.01$, and
the leading oscillation scale $\delta m_{32}^2=3.5\times 10^{-3}$~eV$^2$. 
The expected $\nu_\mu \rightarrow \nu_e$ signal is compared in 
Fig.~\ref{sb} (left panel) to various background components. 
In particular, the NC background spectrum is shown after requiring
that the electron--candidate energy is at least 30\% of the total energy,
which suppresses the $\pi^0$'s. 
Note that the surviving 
NC background is still the dominant background component, 
although the others are not negligible. 
The total background level is about equal to the 
signal level for this particular example. 
The signal/background ratio could be improved by narrowing the 
energy spread of the neutrino beam. For example, if neutrinos with 
energies between 8~GeV and 10~GeV are selected, 
the right--hand panel in Fig.~\ref{sb} shows the resulting signal and 
background distributions. Note that for this case the NC background level
is roughly 1/6 of the signal.

In summary, by requiring the electron candidate to have a significant 
fraction of the visible event energy, the NC background can be reduced so 
that its contribution to $f_B \sim 0.04$, at the price of reducing $D$ 
by $\sim 30$\%. 
Further reductions in the NC background depend on detector 
details, and will be discussed in the following sub--sections.

%\subsection{Detectors and further NC background rejection} 
%
%In the following we will consider liquid argon, water cerenkov, 
%and iron/scintillator neutrino detectors. These three detector--types 
%have very different capabilities for further NC background rejection. 
%
%Recall that in the muon 
%storage ring beam the most important aspect of any detector was 
%the signal efficiency and the appearance of wrong-sign muons from 
%hadronic showers, which varied at most by about a factor of 2 or 3 
%between detectors (cite).  For the conventional beams the different 
%detectors have background rejection factors which can vary by an order
%of magnitude.  At the end of this section we will discuss briefly what 
%the cost considerations are for the different detectors, as well as 
%summarize the NC background level expected in each one.  

\subsubsection{Liquid Argon TPC} 

In a liquid argon TPC 
there are four ways that NC events can contribute backgrounds
in an electron appearance search: 
a) a photon converts very close to the interaction vertex, b) a charged 
pion interacts very close to the interaction vertex, c) a charged pion
overlaps with a $\pi^0$ creating an electron-like track, or 
d) there is an asymmetric Dalitz decay or external conversion of a $\pi^0$ 
created at the interaction vertex.  
The imaging capability of a liquid argon TPC of the type 
planned by the ICARUS collaboration is expected to facilitate 
much better rejection of these backgrounds than any other high--mass 
neutrino detector that exists or is planned. 

\begin{figure}
\epsfxsize=.5\textwidth 
%\centerline{\epsffile[90 440 400 700]{el_dal.ps} }
\centerline{\epsffile[90 440 400 700]{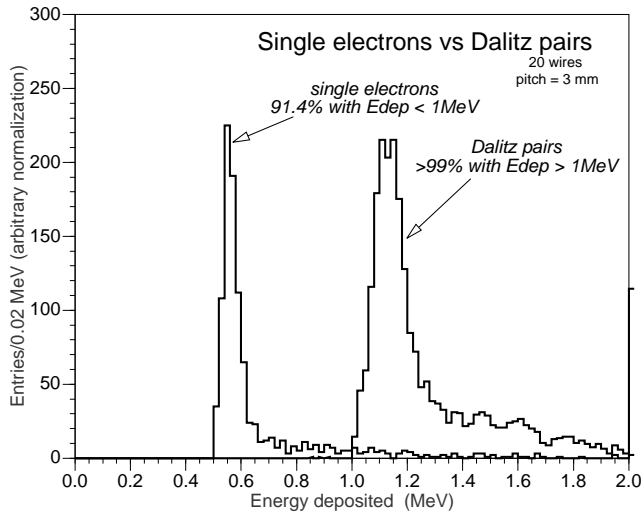} }
\caption{Simulated dE/dx measurement in liquid argon for both 
single electrons and Dalitz pairs based on 
summing over 20 wires in the ICARUS detector, or 6~cm 
of Liquid Argon.}
\label{fig:dalitz}
\end{figure}

A full simulation of the ICARUS detector~\cite{icarus} in the 
CNGS beam~\cite{cngs} 
(which has a mean neutrino energy of about 17~GeV) shows that 
background sources a) and d) have the largest rates, and 
contribute respectively 4262 and 275 out of 15550 NC events. 
Note that  there are two charged particles resulting from 
the Dalitz decay or the photon conversion. 
These backgrounds can therefore be suppressed 
by requiring that the electron candidate does not have an unusually 
high dE/dx before it showers. 
Figure~\ref{fig:dalitz} shows an ICARUS Monte Carlo
simulation of the electron/Dalitz pair separation using the first 6~cm 
of the ``track'' after the interaction vertex.  
Requiring the deposited energy to be less than 1~MeV would keep 90\% 
of the signal while removing over 99\% of 
the two-electron background. 
Requiring the reconstructed electron energy to exceed 1~GeV would 
further reduce these backgrounds by about a factor of 3. 
Hence, before any other kinematic or reconstruction cuts are applied, 
backgrounds a) and d) can be reduced to $\sim 10^{-3}$ of 
the NC event rate~\cite{icanoetdr}.  

The NC background sources b) and c), where a single charged 
pion either interacts or 
overlaps with a $\pi^0$, have a low rate (31.5/15550 NC events).  Once 
again, two 
thirds of these events can be removed by requiring that the electron candidate 
has an energy greater than 1~GeV. Hence, backgrounds b) and c) 
can be reduced to less than $10^{-3}$ of the NC event rate, at the cost of 
reducing $D$ by a factor of $\sim 0.9$.

We conclude that, in an ICARUS--type detector, 
the total contribution to the background fraction $f_B$ from NC events 
can be reduced to $\sim 0.001$, at the cost of a modest reduction in 
the dataset size $D$ of $\sim10-15$\%.

\subsubsection{Water Cerenkov} 

Although Super-Kamiokande can easily discriminate between quasi-elastic
$\nu_e$ and quasi-elastic $\nu_\mu$ events, this becomes problematic in the 
presence of a hadronic shower with a significant number of high energy 
secondaries. It was shown~\cite{jhfloi} that for neutrino energies 
of $\sim 1$~GeV, the NC background contribution to $f_B$ 
can be reduced to $\sim 0.03$ at the cost of a reduction in $D$ 
by a factor of 0.68. 
We will assume that, with further optimization, $f_B$ can be reduced 
to $\sim 0.02$. 
The $\pi^0$'s in NC events are rejected by exploiting 
the cerenkov ring structure (two separated rings from 
$\pi^0 \to \gamma\gamma$ decays, one ring for an electron shower). However,  
above neutrino energies of $\sim 1$~GeV the two photons
from the $\pi^0$ decay are indistinguishable, and the only  
discrimination against NC backgrounds is based on 
the electromagnetic energy fraction in the event~\cite{dcasper}. 
Hence we conclude that for 1~GeV neutrino beams the 
contribution to $f_B$ from NC events can be reduced to about 0.02. 
For higher energy beams the backgrounds are much larger, and we will 
assume that $f_B = 0.04$. 
%in which case a 
%$\nu_\mu \rightarrow \nu_e$ search with an interesting level 
%of sensitivity is probably precluded. 

\subsubsection{Sampling Calorimeters}

Sampling calorimeter neutrino detectors typically consist of a sandwich 
of target material (Iron, glass, Lead)
and active material (scintillator or wire chambers). 
Relatively crude sampling calorimeters offer a compact and cost effective and 
way of instrumenting a large fiducial mass for muon detection, 
and hence $\nu_\mu$ CC measurements.  
However, a $\nu_\mu \rightarrow \nu_e$ experiment requires 
an electron appearance measurement, which 
imposes significant additional
requirements on the granularity of the calorimeter. 
In order to distinguish the 
electron produced in a neutrino interaction from $\pi^0$ production,
the calorimeter must be finely segmented both longitudinally and
transversely. The longitudinal segmentation is needed to resolve the
difference between 1 particle (an electron) and 2 particles (from
a photon conversion) at the beginning of the electromagnetic
shower. The transverse segmentation is needed to prevent accidental
overlaps of charged particles which would veto a real electron. 
We consider both fine--grained and course--grained sampling calorimeters:
\begin{description}
\item{(i) Fine grained calorimeter.} 
The Soudan 2 detector is an example of a finely segmented iron
calorimeter with drift tube readout. The steel plates are 3.2~mm
(0.18~$X_0$) thick. The estimated fraction of NC 
events which fake $\nu_e$ interactions is 2.3\% after applying 
cuts that include requiring 
the candidate electron energy to be at least 50\% of the visible energy. 
Hence, the contribution to $f_B$ from NC events in a Soudan~2 type detector 
is $\sim0.01$. The associated reduction in $D$ is by a factor of 0.6.
\item{(ii) Course grained calorimeter.} 
The MINOS detector is an example of a course grained iron calorimeter 
with scintillator readout. The toroidally magnetized steel plates 
are 1~inch thick, and the detector is instrumented with 4.1~cm wide 
scintillator strips. NC background events can be rejected based on 
the electron candidate energy, the total event energy, and the number of 
scintillator strips associated with the electron shower. A recent 
study~\cite{wai} suggests that with the medium energy NuMI beam, the 
NC backgrounds for a $\nu_\mu \to \nu_e$ search in a MINOS--like 
detector can be reduced so that the contribution to $f_B$ is $\sim 0.01$, 
at the cost of reducing $D$ by a factor of 0.33.
\end{description}

\subsection{Charged Current Backgrounds}

In general, in a $\nu_\mu \rightarrow \nu_e$ search most $\nu_\mu$ CC events 
can be rejected due to the presence of a muon. In our present 
study  we only consider the $\pi^0$ backgrounds from NC events.
However, for the NuMI ME beam 
our simulations show that if all muons below 2~GeV
in CC events are missed, the $\pi^0$ background from CC events will roughly be
equal to the corresponding background from NC events. 
The muon detection threshold is detector dependent, 
and more detailed detector studies are needed for the CC background to be  
taken into account. 

\subsection{ Backgrounds from $\nu_\mu \to \nu_\tau$ oscillations} 

\begin{figure} 
\centerline{
\epsfxsize=.5\textwidth\epsfbox{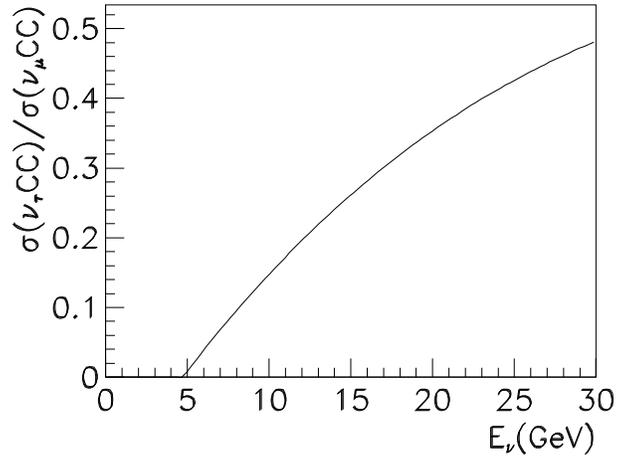} }
\caption{The $\nu_\tau / \nu_\mu$ CC 
cross section ratio as a function of neutrino energy~\cite{maury}.
}
\label{fig:tau1} 
\end{figure}

The oscillation process $\nu_\mu \to \nu_\tau$ can produce a 
$\nu_\mu \to \nu_e$ 
background if the $\nu_\tau$ interacts to produce a tau--lepton that 
subsequently decays electronically ($\tau \to e + X$, $BR = 20$\%).  
If the detector does not have good discrimination between electrons and 
neutral pions, the process $\tau\to n\pi^0 X \nu_\tau$ ($BR = 37$\%) also 
contributes to the background.  The $\nu_\tau$ backgrounds are particularly 
dangerous because they have the same dependence on 
$\delta m_{32}^2 L/E$ as the 
signal.
%, and therefore the signal/background ratio from this source 
%does not decrease with increasing $L$. 
Figure~\ref{fig:tau1} shows the $\nu_\tau / \nu_\mu$ CC 
cross section ratio as a function of neutrino energy~\cite{maury}. 
Clearly the $\nu_\tau$ backgrounds can be 
eliminated by running below, or near to, the $\nu_\tau$ 
CC threshold (5~GeV). 

\begin{figure} 
\epsfxsize=\textwidth\epsfbox{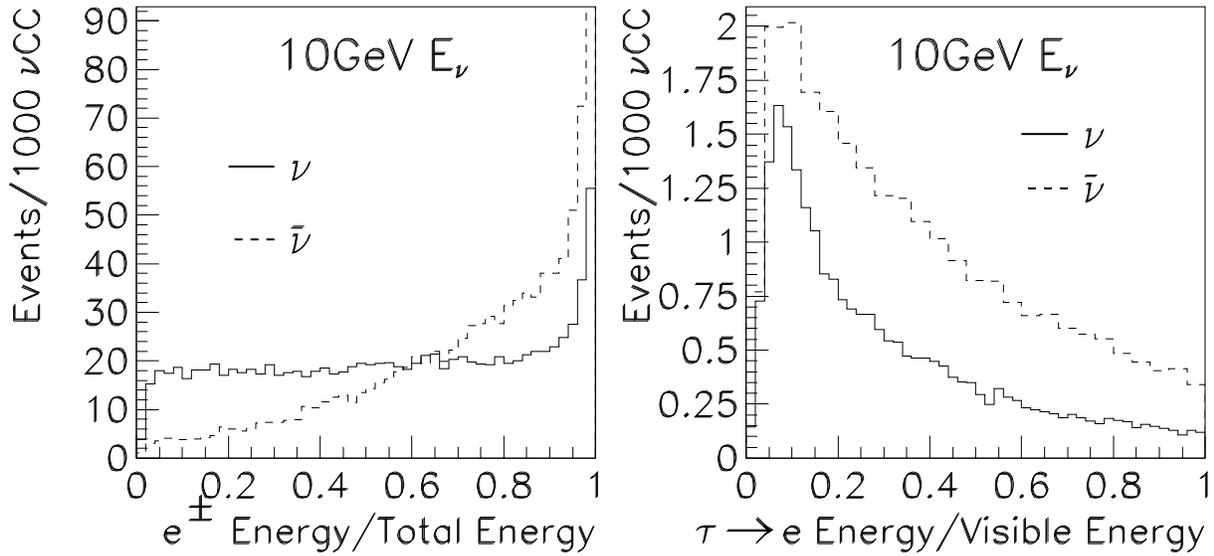} 
\caption{Ratio of electron energy to total visible energy for (left)  
 $\nu_e$ and $\bar\nu_e$ charged current events and (right) 
 $\nu_\tau$ and $\bar\nu_\tau$ charged current events, 
where the $\tau$ subsequently decays to an electron.} 
\label{fig:tau2} 
\end{figure} 
\begin{figure}
\epsfxsize=.99\textwidth
%\epsfbox{fbtau.eps}
\epsfbox{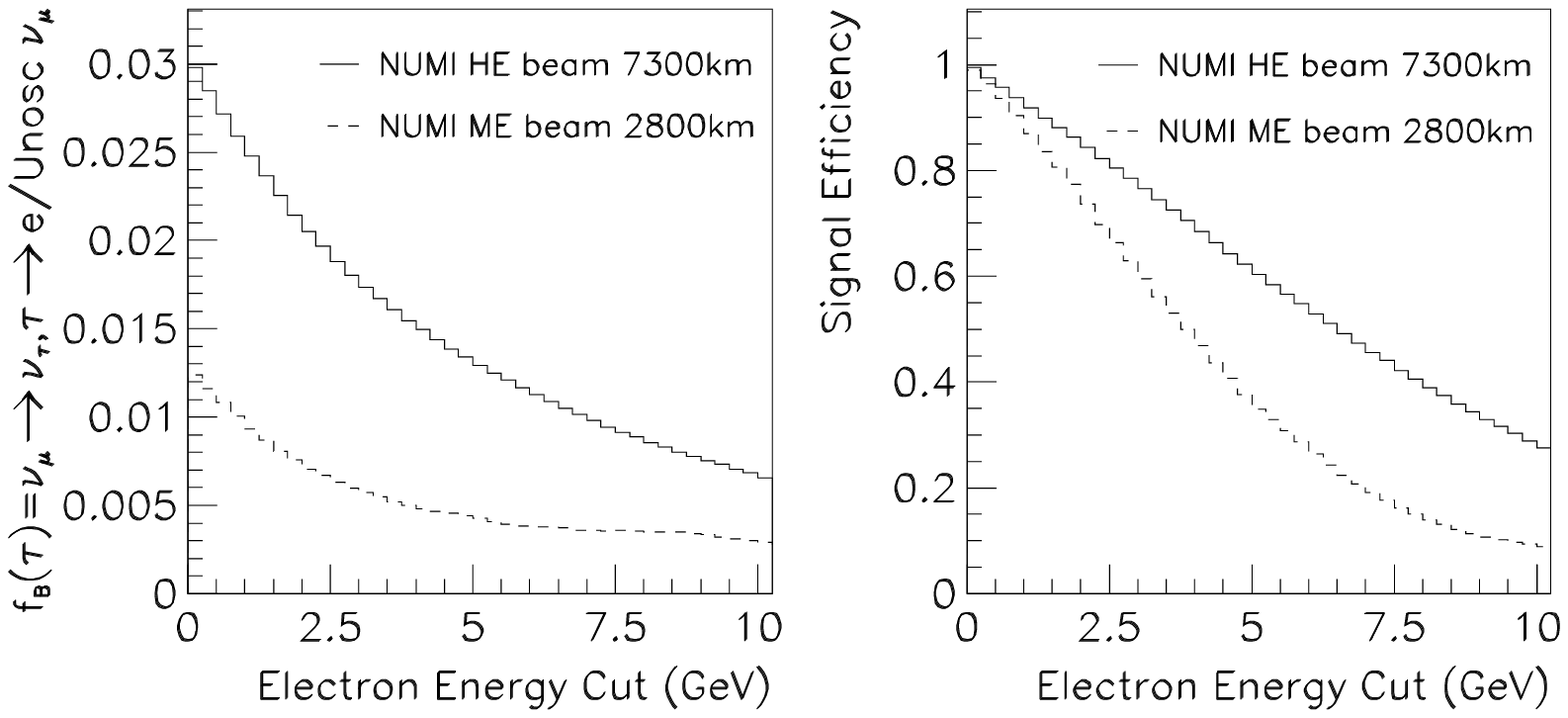} 
\caption{ For oscillations with $\delta m^2_{32} = 3.5 \times 10^{-3}$~eV$^2$ 
and $\sin^2 2\theta_{23} = 1$, the $\nu_\mu \to \nu_e$ background contribution 
$f_B(\tau)$ arising $\nu_\tau$ interactions followed by $\tau \to e$ decays 
(Left panel), and the associated $\nu_\mu \to \nu_e$ 
signal efficiency (right panel) shown 
as a function of a cut on the minimum electron energy.
The energy resolution assumed
for electrons is $\sigma(E_e)/E_e = 5\%/\sqrt{E_e(GeV)}$.
}
\label{fig:fbt}
\end{figure}

If the neutrino beam energy is significantly above 5~GeV, 
one way to reduce the $\nu_\tau$ background is to require 
that the electron candidate carries a significant fraction of the 
total observed interaction energy. 
Assuming a perfect detector energy resolution, the distributions of 
electron energy divided by total energy are shown in 
Fig.~\ref{fig:tau2} for 10~GeV $\nu_e$ and $\bar\nu_e$ CC events (left), 
and 10~GeV $\nu_\tau$ and $\bar\nu_\tau$ CC events (right) 
where the $\tau$ subsequently decays to an electron.   
%The plots are normalized to the same number of incoming neutrinos. 
If the electron candidates are required to carry at least $50\%$ 
of the visible energy, the remaining $\nu_\tau$ background level 
(for $\sin^2(1.27 \delta m_{32}^2L/E)=1$)  
is a few per cent at 10~GeV. 
Note that the backgrounds for $\nu$ and $\bar\nu$ running are different.  
Better background rejection (for a given reduction in signal efficiency) 
can be obtained by imposing a minimum energy requirement on the 
electron candidate instead of a requirement on the fraction of the 
total energy carried by the electron.  
The contribution to $f_B$ from the $\tau \to e + X$ backgrounds is shown in 
Fig.~\ref{fig:fbt} as a function 
of the minimum electron energy requirement (left panel) for the NuMI 
high energy beam with $L = 7300$~km, and for the medium energy beam 
with $L = 2800$~km. The corresponding reductions in $D$ are shown in 
the right panel of Fig.~\ref{fig:fbt}. 
Table~\ref{tab:tausum} summarizes the expected contribution to 
$f_B$ for three NuMI beams using some explicit electron energy cuts.
For the medium-- and high--energy beams, to reduce the contribution to 
$f_B$ to $\lsim 0.01$ requires a significant reduction in $D$. 
Furthermore, the total contribution to $f_B$ from $\nu_\tau$ interactions 
can be a factor of $\sim 2$ larger than listed in Table~\ref{tab:tausum}
if the detector does not provide adequate discrimination against 
$\tau\to n\pi^0 X \nu_\tau$ decays.   

\begin{table}
\caption{Summary of $\nu_\tau$ backgrounds in which $\tau \to e + X$. 
The background rates are listed for NuMI--like low--, 
medium--, and high--energy beams, assuming the oscillation parameters 
$|\delta m_{32}^2| = 3.5\times 10^{-3}eV^2$, 
and $\sin^2 2\theta_{23}=1$.  The assumed electron energy resolution 
is $\sigma(E_e)/E_e = 5\%/\sqrt{E_e(GeV)}$, as anticipated for the 
ICARUS detector. 
Note that the total $\nu_\tau$ related backgrounds can be a factor 
of $\sim2$ larger if the detector does not have good descrimination 
against $\tau\to n\pi^0 X \nu_\tau$ decays.}
\centerline{ 
\begin{tabular}{lccccc}
&&&&&\\
\hline  
Beam & Distance & $f_B(\tau)$ & $f_B(\tau)$ & 
$E_e$ & Signal \\
 & (km) &no cut&with cut& cut&Efficiency\\
\hline
NuMI-LE & 732   & 0.001 & 0.001 & none   &  1 \\
NuMI-ME & 2800  & 0.013 & 0.006 & $>3$ GeV   &  0.6 \\
NuMI-HE & 7300  & 0.031 & 0.013 & $>5$ GeV   &  0.6 \\
\hline
\end{tabular}
}
\label{tab:tausum}  
\end{table} 
\begin{figure}
\centerline{
\epsfxsize=.45\textwidth\epsfbox{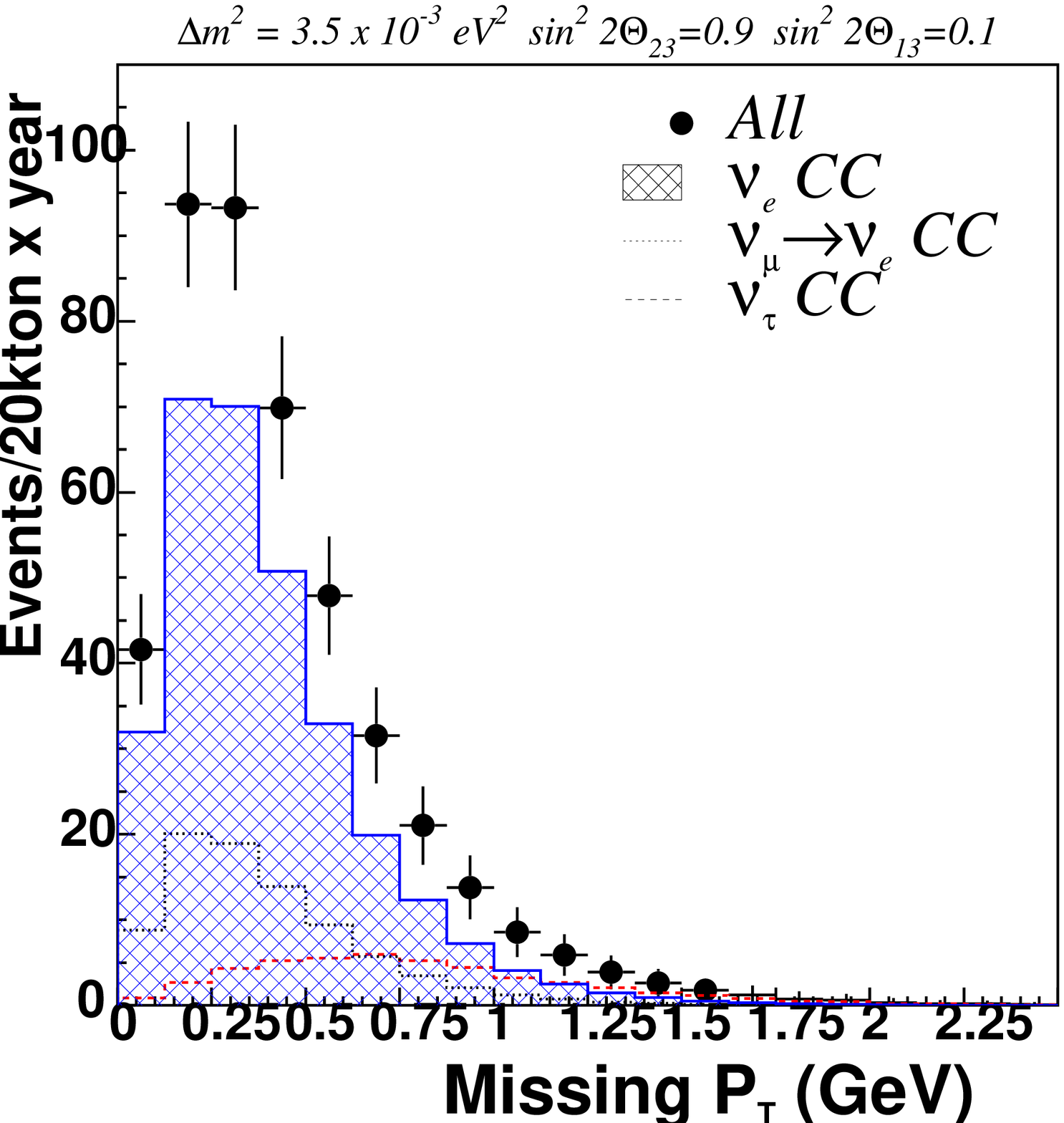} 
\epsfxsize=.45\textwidth\epsfbox{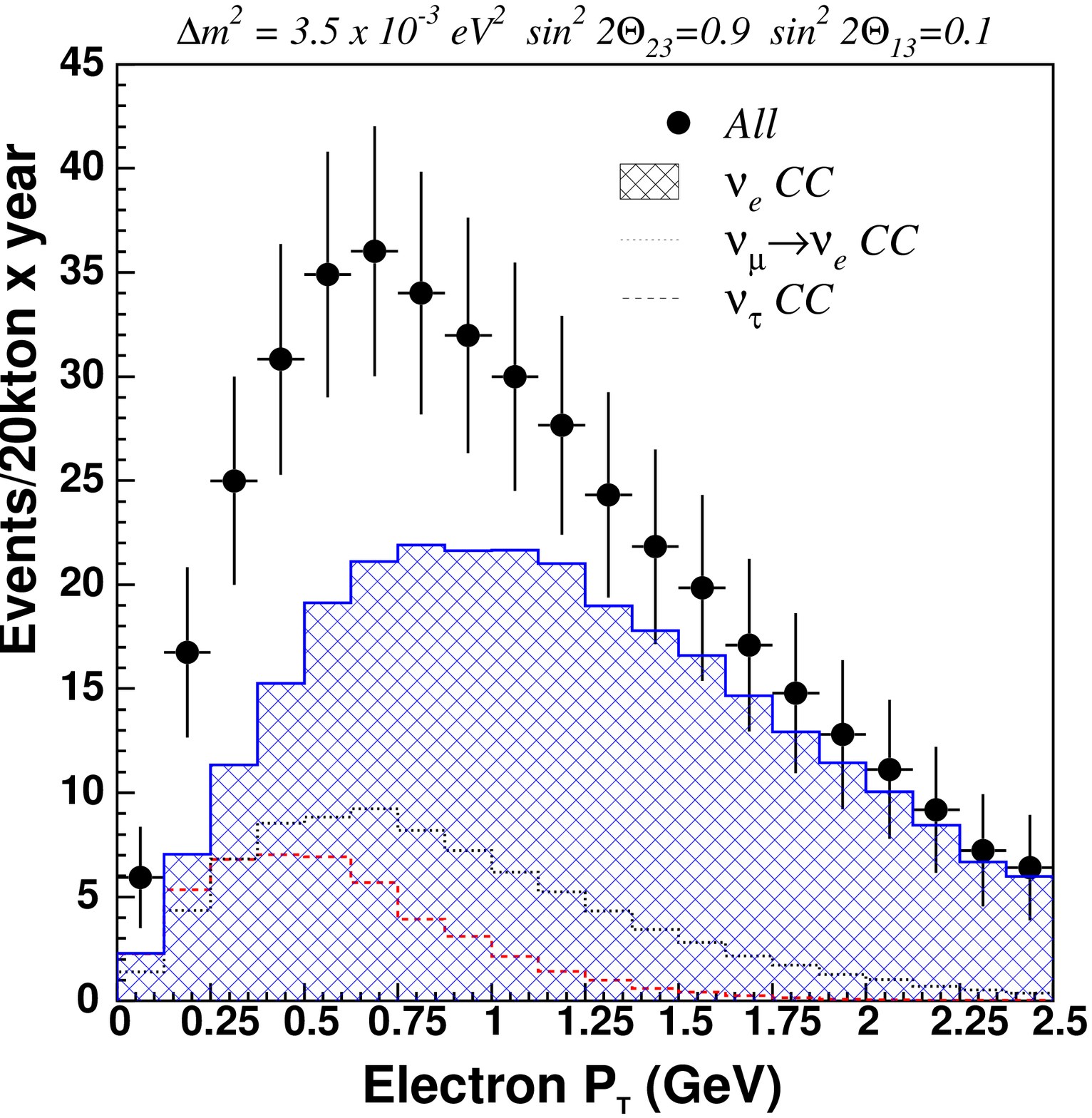} 
}
\caption{Missing transverse momentum (left) and electron transverse 
momentum (right) distributions for $\nu_e$ charged current
events and for $\nu_\tau, \tau\to e$ events in the ICARUS detector, 
assuming oscillation parameters as indicated.}
\label{fig:icanoetau} 
\end{figure} 

There are other kinematic handles that can in principle be used to 
suppress $\nu_\tau$ backgrounds, but the performance of a set of 
given kinematic cuts depends upon the detector details. 
Obvious kinematic quantities that can be exploited are 
the missing transverse momentum (penalizing the missing energy 
associated with the neutrinos produced in the $\tau$--lepton decay), 
and the electron transverse momentum distribution.  
The simulated distributions for these variables 
are shown~\cite{icanoe2} in Fig.~\ref{fig:icanoetau} 
for several event types in the ICARUS detector at the CNGS beam. 
Note that the CNGS beamline has a $\nu_e$ background 
fraction of 0.8\% and a mean neutrino energy of 17~GeV, and  
$\sin^2 2\theta_{13}$ has been assumed to be 0.1, just at the current limit. 
The distributions shown in the figure are for all electron-like 
events with no electron energy requirement. 
It should be noted that events that pass an electron energy 
cut will tend to have electrons moving 
close to the $\nu_\tau$ direction, and hence are more likely to survive 
other kinematic cuts.
For an experiment in which the $\tau$ appearance rate is 
likely to be significant, the ICARUS collaboration finds the best sensitivity
is obtained with a combined fit, rather than by eliminating the $\nu_\tau$ 
events.

We conclude that $\nu_\tau$ backgrounds are only significant for the 
medium-- and high--energy beams, in which case an electron energy cut 
and/or good $\tau \to e$ 
rejection is needed to reduce the contribution to $f_B$ to $\lsim 0.01$.
This might be accomplished in a liquid argon detector, which provides 
good discrimination against $\tau\to n\pi^0 X \nu_\tau$ decays. 
The $\tau$ related backgrounds might be larger in other detectors. 
For example, a recent study~\cite{wai}  
concluded that, with $\delta m^2_{32} = 0.003$~eV$^2$ and $L = 2900$~km, 
after cuts which reduced $D$ by a factor of 0.33, 
the contribution to $f_B$ from $\nu_\mu \to \nu_\tau$ associated 
interactions was 0.02 for a MINOS--like iron--scintillator detector at 
a NuMI--like medium energy beam.

\subsection{ The systematic uncertainty on the background } 

Since a $\nu_\mu \to \nu_e$ search will not be background free, 
the background must be subtracted from any potential signal. 
The background subtraction introduces a systematic uncertainty 
associated with the imperfect knowledge of the expected background rate.  
Assigning a systematic error on the background 
rate requires understanding the uncertainty on the 
$\nu_e$ contamination in the beamline, and the uncertainty on
the detector's ability to reject background events.

To understand the $\nu_e$ fraction in the initial neutrino beam, we 
must know the charged and neutral kaon components in the 
secondary beam. This requires knowledge of their production 
at the proton target, and knowledge of the beamline acceptance. 
In addition, the 
backgrounds from muons decaying in the decay tunnel must be understood, 
although generally
speaking the uncertainty from this background component is relatively small.
Measuring the fraction of charged and neutral kaons produced in
the target is the subject of much current study. The experiment 
P907~\cite{p907} is being
proposed at Fermilab to measure the kaon production cross sections 
for 120~GeV protons on the MINOS target. A
similar experiment, HARP~\cite{harp}, 
is planned at CERN to study meson production 
for the CNGS beam. P907 and HARP are expected to significantly 
reduce the kaon production uncertainties for neutrino beams using 
120~GeV and multi--GeV (up to 15~GeV) proton beams, together with targets 
similar to NuMI and CNGS. However, the combined effects of both
production and acceptance must be understood. 
For this purpose, a very fine-grained near 
detector could be used.  As an example, the NOMAD detector~\cite{nomad} can 
determine 
the neutral kaon contribution to the $\nu_e$ flux by comparing the 
rates of $\nu_e$ and $\bar\nu_e$ events that are seen. 
With an electron charge analysis to discriminate between 
$\nu_e$ and $\bar\nu_e$ events, the NOMAD experiment is able to limit the
systematic uncertainty on the overall near detector 
$\nu_e$ flux to 2\%, and the
energy spectrum shape to 2.5\%~\cite{cousins}.  With 
a fine-grained near detector of this type one might imagine 
achieving a systematic uncertainty on the far detector flux of 3 to 4\%.

Background rejection in the far detector can be understood using a 
second near detector of the same type as the far detector, 
Assuming that appropriate near detectors are used, 
the remaining systematic
uncertainties on the backgrounds in the far detector come from 
the slightly different beam spectrum that any near detector sees, 
and the uncertainties on the differences in fiducial volumes and event 
acceptances.  
While there is probably no hard limit on the systematic uncertainty 
that could be achieved, it is unreasonable to expect that the total systematic
uncertainty on the background fraction could be reduced below a few per cent of
the background itself.  
In the following we will assume that the uncertainty on the background rates 
is 5\%. Note that the participants of a recent study of 1~GeV neutrino beams 
at the JHF~\cite{jhfloi} assumed the more conservative value of 10\% for the 
systematic uncertainty on the backgrounds.

\subsection{Summary: Dataset size $D$ \& background fraction $f_B$}

In the previous discussion we concluded that, 
for a $\nu_\mu \to \nu_e$ measurement, 
the contribution to the expected background 
fraction $f_B$ from the initial $\nu_e$ beam contamination might be reduced 
to $\sim 0.002$. However, 
for most of the detector types we have considered, $f_B$ is dominated by the 
contributions from 
(a) neutral pions faking an electron signature, and/or 
(b) $\nu_\mu \to \nu_\tau$ related backgrounds. We would 
like to know, as a function of detector choice, the values of $f_B$ and 
$D$ that should be used in assessing the $\nu_\mu \to \nu_e$ physics potential.
The dominant contributions to $f_B$, together with $f_B$ and the 
associated signal efficiencies, 
and detector cost estimates, are summarized in  
Table \ref{tab:bksum}, along with the implied value of $D$ for 5~years of 
data taking. The values of $D$ are estimated assuming the detectors cost 
\$500M, which determines the detector masses. 
Estimating the unit costs for each detector 
type is not straightforward. Details of the cost estimates are given 
in Appendix~2. 

\begin{table}
\caption{Detector background rates ($f_B$), signal efficiencies, 
and unit costs. Water cerenkov backgrounds and 
efficiencies are neutrino energy dependent: numbers left 
of the arrows for a 1~GeV beam, numbers right of the arrows for 
a multi--GeV beam requiring $y < 0.5$.} 
\begin{tabular}{lccccc} 
&&&&&\\
\hline
    & Water Cerenkov & Liquid Argon & \multicolumn{2}{c}{Steel+readout}& 
Liquid$^{f}$ \\
    &   (UNO)    & (ICARUS)     & (MINOS) & (THESEUS) &Scintillator \\
\hline
Signal Efficiency    & 0.7$\to$ 0.5   &  0.90   & 0.33 & 0.6$^b,g$  & 0.76  \\
$f_B$(NC)   &$0.02\to 0.04$& 0.001  & 0.01 & 0.01  &   $<0.006$    \\
$f_B$(beam) & 0.002        & 0.002  & 0.002& 0.002 &  0.002    \\
$f_B(\tau)$&$0 \to 0.01$ & $\sim 0.005^c$  & $0.02^c$&
$0^b$  & $\sim 0.005^c$ \\ 
$f_B$     &$0.02\to 0.05$& $\sim 0.008$  & $0.03$ & 0.01  &  $\sim 0.01$   \\
Electron cut   &  $>0.5\times E_\nu$ & none$^d$  
& 1-6 GeV & $> 0.5 E_{vis}$ & $E_{vis} > 2$ GeV     \\
Unit cost (M\$/kt)$^a$ & 2.4 &  23  & 10.4 & 78 & 59 \\
Mass (kt) / 500~M\$   & 745 & 37 & 85 &  6.4  & 260  \\
$D$ (kt-yrs)$^e$   & 2600 $\to$ 1860 & 170 & 140 & 19 & 990  \\
\hline \\
\end{tabular}
\\\\
$^a$ FY00 dollars. Costs account for salaries, overheads, and contingencies. 
Details are given in Appendix~2. 
The cost does not include excavating a cavern, which is  
estimated~\cite{uno} to be 0.5M\$/kton/$\rho$,  ($\rho =$ target density).\\
$^b$ For the MINOS low energy beam.\\
$^c$ For the MINOS medium energy beam.\\
$^d$ Although a total energy cut might be applied to reduce the intrinsic 
$\nu_e$ background.  \\
$^e$ For 5~years running. \\
$^{f}$ $\nu_e n \to e^- p$ search, Ahrens et al, Phys. Rev. D volume 36, 
(702) 1987. \\
$^g$ Soudan efficiency for electron neutrinos: NuMI-L-562.
\label{tab:bksum}
\end{table} 

Note that no detector achieves the goal 
$f_B < 0.004$ that we derived in Section~4 for a multi--GeV beam. Once 
above the $\nu_\tau$ CC threshold the contribution to $f_B$ from $f_B(\tau)$ 
already exceeds 0.004. As an example, consider the 
liquid Argon detector. The parameters to be used for 
this detector type with a medium energy MINOS--type beam are: 
$D = 170$~kt-yrs, $f_B = 0.008$, 
and $\sigma_{f_B}/f_B = 0.05$ (as discussed in Section~5.5).
If the $\nu_\tau$ background contribution could be eliminated 
$f_B$ would be reduced to $\sim 0.003$, but this improvement in 
background rate would be accompanied by a significant decrease in $D$, and 
would not be expected to significantly improve the 
sensitivity to $\nu_\mu \to \nu_e$ oscillations.

\section{ Physics with multi--GeV long--baseline beams } 

Using the values of $D$ and $f_B$ in Table~\ref{tab:bksum}, and assuming 
$\sigma_{f_B}/f_B = 0.05$, we can now assess the physics potential for the 
various detector scenarios we have considered. We will begin with the 
multi--GeV long baseline beams, and consider the 
minimum value of $\sin^2 2\theta_{13}$ that will yield a $\nu_\mu \to \nu_e$ 
signal $3\sigma$ above the background, the sensitivity 
to the neutrino mass hierarchy, and the sensitivity to CP violation in the 
lepton sector.

\subsection{$\sin^2 2\theta_{13}$ Reach } 

To obtain the $\sin^2 2\theta_{13}$ reach for the detector scenarios listed 
in Table~\ref{tab:bksum} we return to Fig.~\ref{fig:contours} which shows 
contours of constant reach in the $(f_B, D)$--plane for upgraded (1.6~MW) 
NuMI medium-- and high--energy beams. Of the scenarios we 
have considered, the greatest sensitivity is obtained using a liquid argon 
detector with either the medium-- or high--energy beams at 
$L = 2900$~km, 4000~km, or 7300~km. 
In these cases  a $\nu_\mu \to \nu_e$ signal 
at least $3\sigma$ above the background would be expected provided 
$\sin^2 2\theta_{13} > 0.002$ to $0.003$. 
If the $\nu_\tau$ backgrounds could be 
elliminated, reducing $f_B$ to 0.003, the limiting sensitivity improves to 
$\sin^2 2\theta_{13} > 0.001$. If the initial $\nu_e$ contamination in 
the beam is 0.5\% (rather than the assumed 0.2\%), so that $f_B = 0.01$, the 
$\sin^2 2\theta_{13}$ reach is still $\sim 0.002$ to $0.003$. 
Hence, the estimated 
reach is not very sensitive to the uncertainties on our  
background estimations. However, the $\nu_\mu \to \nu_e$ sensitivity 
would be degraded if the $\nu_\tau$ backgrounds were significantly larger, 
which disfavors higher beam energies.

We conclude that, with a 30-40~kt liquid argon detector and a medium 
energy superbeam, we could improve the sensitivity to $\nu_\mu \to \nu_e$ 
oscillations, and obtain about an order of magnitude improvement in 
the $\sin^2 2\theta_{13}$ reach beyond that expected for 
the currently approved 
next generation experiments. The other detector choices in 
Table~\ref{tab:bksum} do not seem to be competitive with liquid argon, 
with the possible exception of the 
water cerenkov detector which obtains a reach of about 0.003 with the 
high energy beam at the longest baseline ($L = 7300$~km).

\begin{figure}
\centering\leavevmode
\epsfxsize=3.7in\epsffile{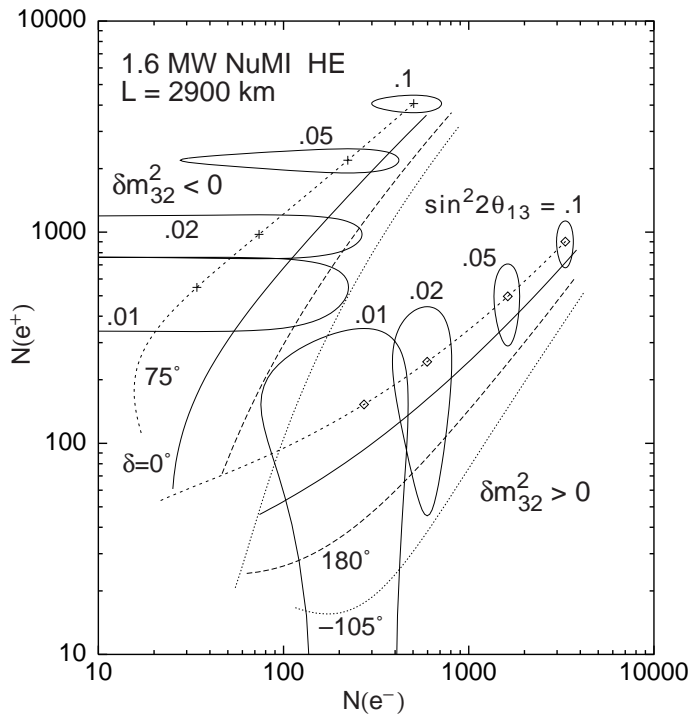}
%\medskip
\caption[]{Three--sigma error ellipses in the 
$\left(N(e^+), N(e^-)\right)$--plane, shown for 
$\nu_\mu \to \nu_e$ and $\bar\nu_\mu \to \bar\nu_e$ oscillations 
in a NuMI--like 
high energy neutrino beam driven by a 1.6~MW proton driver. 
The calculation assumes a liquid argon detector with the parameters 
listed in Table~\ref{tab:bksum}, a baseline of 2900~km, 
and 3~years of running with neutrinos, 6~years running 
with antineutrinos. 
Curves are shown for different CP--phases $\delta$ (as labelled), and 
for both signs of $\delta m^2_{32}$ with 
$|\delta m^2_{32}| = 0.0035$~eV$^2$, and 
the sub--leading scale $\delta m^2_{21} = 10^{-4}$~eV$^2$. 
Note that $\sin^22\theta_{13}$ varies along the curves from
0.0001 to 0.01, as indicated~\cite{geersb}.
}
\label{fig:signdm2}
\end{figure}
\begin{figure}
\centering\leavevmode
\epsfxsize=3.1in\epsffile{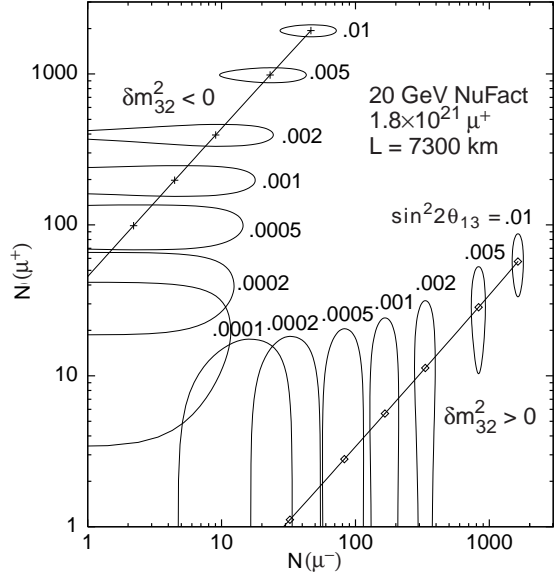}
%\medskip
\caption[]{Three--sigma error ellipses in the 
$\left(N(\mu+), N(\mu-)\right)$--plane, shown for a 20~GeV neutrino 
factory delivering $3.6\times10^{21}$ useful muon decays and
$1.8\times10^{21}$ antimuon decays, with a 50~kt
detector at $L = 7300$~km, $\delta m^2_{21} = 10^{-4}$~eV$^2$, 
and $\delta = 0$. Curves are shown for both signs of
$\delta m^2_{32}$; $\sin^22\theta_{13}$ varies along the curves from
0.0001 to 0.01, as indicated~\cite{geersb}.
}
\label{fig:nufact}
\end{figure}

\subsection{Matter Effects and CP Violation} 

Having either established or excluded a $\nu_\mu \to \nu_e$ signal, 
a search for $\bar\nu_\mu \to \bar\nu_e$ over a long baseline can 
determine the sign of $\delta m^2_{32}$, and hence the neutrino mass 
hierarchy. Suppose that $N_+$ and $N_-$ signal events are measured 
in neutrino and antineutrino running respectively. In the absence 
of intrinsic CP--violation, and in the absence of matter effects, 
after correcting for different beam fluxes, experimental livetimes, 
and the neutrino/antineutrino cross section ratio, we would 
expect $N_+ = N_-$. However, if $\delta m^2_{32} > 0$, matter effects 
can reduce $N_+$ and enhance $N_-$. Alternatively, if 
$\delta m^2_{32} < 0$, matter effects can reduce $N_-$ and enhance $N_+$.
For detector scenarios and multi--GeV beams similar to those considered in 
this report, it has been shown~\cite{geersb} that:
\begin{description}
\item{(i)} At $L = 732$~km the expected changes of $N_+$ and $N_-$ due 
to matter effects are modest, and are comparable to changes that might arise 
with maximal CP--violation in the lepton sector. It is therefore difficult to 
observe and disentangle matter from CP effects unless the baseline is longer.
\item{(ii)} At baselines of $\sim 3000$~km or greater matter effects are 
much larger 
than CP effects, and the determination of the sign of $\delta m^2_{32}$ is 
straightforward provided both $N_+$ and $N_-$ have been measured with 
comparable sensitivities, and at least one of them is non--zero. 
The sign of $\delta m^2_{32}$ can be determined with a significance of at 
least $3\sigma$ provided $\sin^2 2\theta_{13} \gsim 0.01$. 
\item{(iii)} CP violation cannot be unambiguously established in any 
of the long baseline scenarios considered.
\end{description}
To illustrate these points we choose one of the most favorable scenarios 
studied: a 1.6~MW NuMI--like high energy beam with $L = 2900$~km, detector 
parameters $f_B$ and $D$ corresponding to the liquid argon scenario in 
Table~\ref{tab:bksum}, and oscillation parameters 
$|\delta m^2_{32}| = 3.5 \times 10^{-3}$~eV$^2$ and 
$\delta m^2_{21} = 1 \times 10^{-4}$~eV$^2$. 
The calculated three--sigma error ellipses in the 
$\left(N(e^+), N(e^-)\right)$--plane are shown in Fig.~\ref{fig:signdm2}
for both signs of $\delta m^2_{32}$, with the curves corresponding to 
various CP--phases $\delta$ (as labelled). The magnitude of 
the $\nu_\mu \to \nu_e$ oscillation amplitude parameter 
$\sin^2 2\theta_{13}$ varies along each curve, as indicated. The 
two groups of curves, which correspond to the two signs of $\delta m^2_{32}$, 
are separated by more than $3\sigma$ provided 
$\sin^2 2\theta_{13} \gsim 0.01$. Hence the mass heirarchy can be determined 
provided the $\nu_\mu \to \nu_e$ oscillation amplitude is not less than an 
order of magnitude below the currently excluded region. Unfortunately, within 
each group of curves, the CP--conserving predictions are separated from the 
maximal CP--violating predictions by at most $3\sigma$. Hence, it will 
be difficult to conclusively establish CP violation in this scenario.

Note for comparison that a very long baseline experiment at a neutrino 
factory would be able to observe $\nu_e \to \nu_\mu$ oscillations and 
determine the sign of $\delta m^2_{32}$ for values of $\sin^2 2\theta_{13}$ 
as small as O(0.0001)~! This is illustrated in Fig.~\ref{fig:nufact}.

\section{Physics with 1~GeV medium baseline beams}  

\begin{figure}
\centerline{
\epsfxsize=.45\textwidth\epsfbox[0 0 520 520]{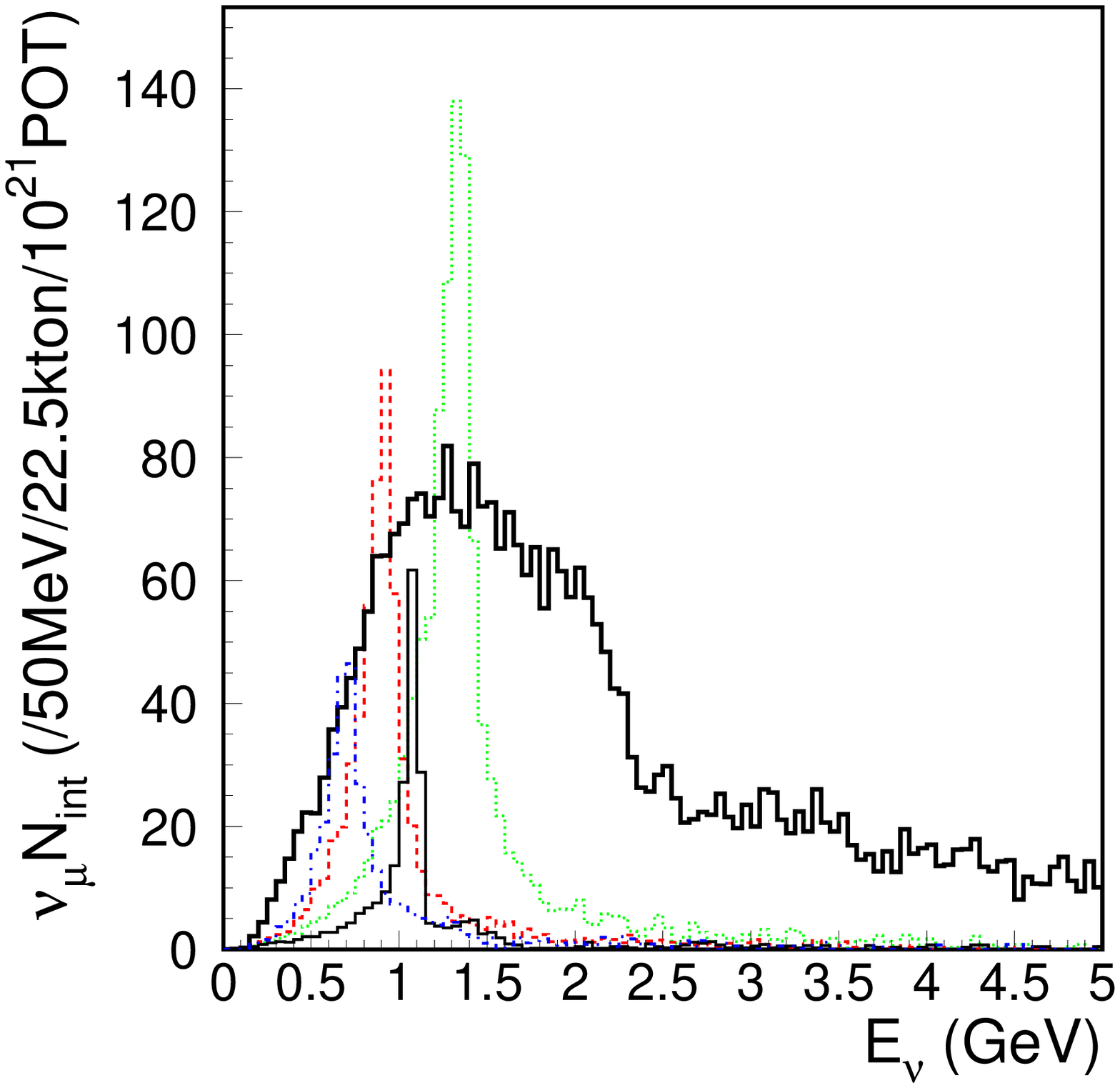}  
\epsfxsize=.45\textwidth\epsfbox[0 0 520 520]{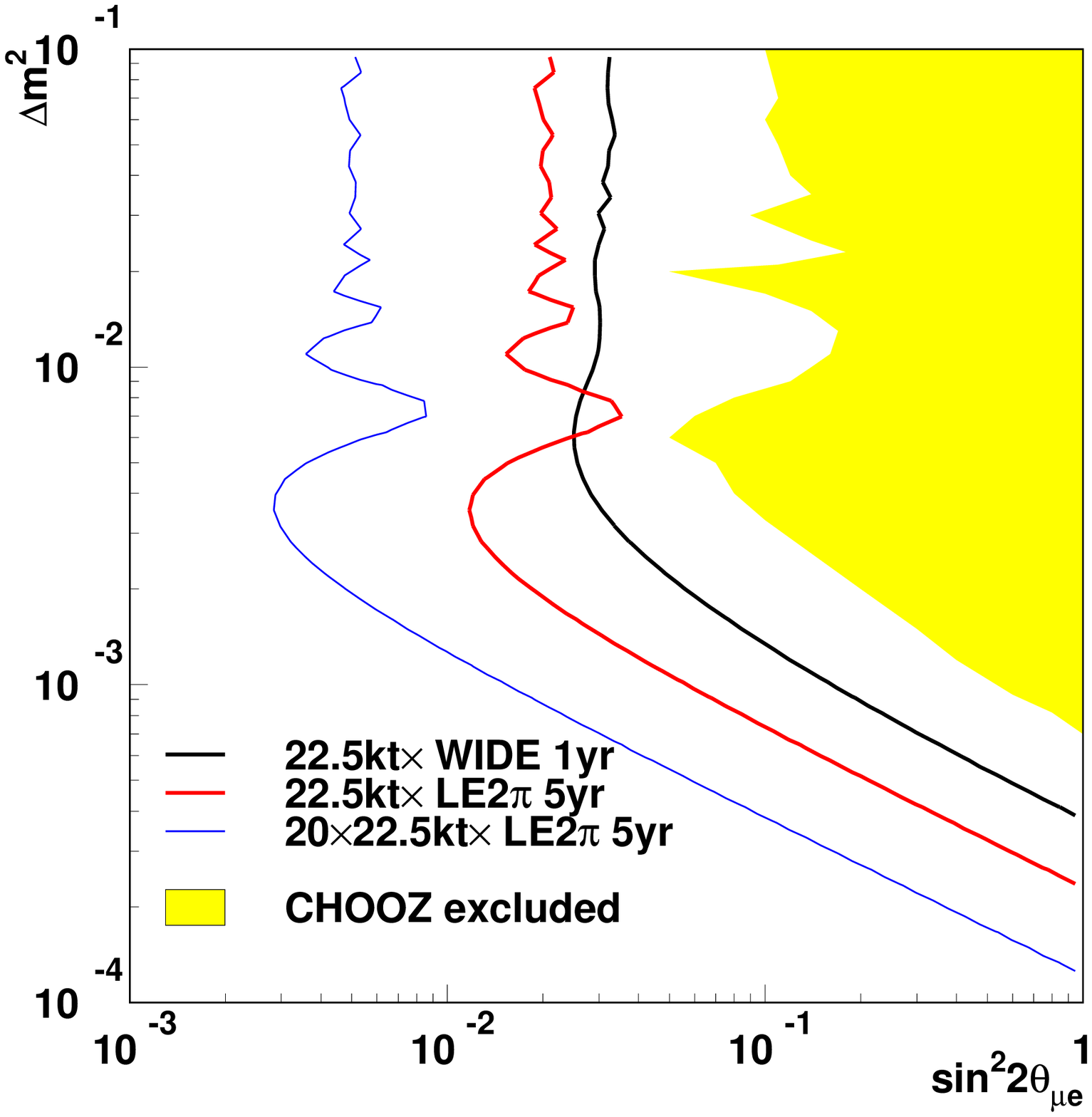} }
\caption{Differential event rates (left) and two--flavor $\nu_\mu \to \nu_e$ 
oscillation sensitivity (right) for a 1~GeV neutrino beam at the JHF 
with $L = 295$~km, and a water cerenkov detector. In the left panel the 
thick solid histogram is for a wide band beam, and the thin solid, 
dashed, dotted, and dot-dash histograms are for four narrow band beam 
designs. The 90\% C.L. contours in the right panel are for, 
as indicated,  1~year with 
a wide band beam, 5~years with a narrow band beam and the SuperK detector, 
and 5~years with a narrow band beam and $20 \times$ the SuperK 
detector~\cite{jhfloi}.}  
\label{fig:jhfloi} 
\end{figure}

We next turn our attention to neutrino beams with energy $E_\nu \sim 1$~GeV. 
The atmospheric neutrino deficit scale $\delta m^2_{32}$ then sets a 
baseline requirement $L \sim 300$~km. 
A recent study~\cite{jhfloi} has generated a letter of interest 
for a 1~GeV neutrino 
beam at the Japan Hadron Facility (0.77~MW 50~GeV proton driver), with 
a baseline of 295~km, using the SuperK 
detector~\cite{superk}.
The JHF study group has considered a variety of beamline designs, including 
both narrow band and wide band beams, quadrupole and horn based focusing. 
The energy distributions for these beams are shown in 
Fig.~\ref{fig:jhfloi} (left panel). 
With a water cerenkov detector the JHF group finds that 
for a $\nu_\mu \to \nu_e$ search it is important to 
use a narrow band beam to avoid the high energy neutrino tail which provides 
a significant source of high energy NC events with detected neutral pions 
that fake lower energy electrons. 
With a narrow band beam, after detailed studies of signal efficiency and 
background rejection they obtain $f_B = 0.03$, dominated by the surviving 
NC backgrounds. This background level was obtained at the cost of reducing $D$ 
by a factor of 0.68. The initial $\nu_e$ component in the beam contributes 
only 0.004 to $f_B$. 
An uncertainty on the background rates of 10\% was assumed.

In the following we discuss the sensitivity of a 1~GeV medium baseline 
neutrino beam to $\nu_\mu \to \nu_e$ oscillations, and the prospects 
for observing CP--violation. Note that baselines of a few hundred kilometers 
are too short for matter effects to be appreciable, and hence the 
pattern of neutrino masses cannot be determined with these medium baseline 
beams.

\begin{figure}
\vspace{-1.0cm}
\centerline{
%\hspace{-1.cm}\epsfxsize=.55\textwidth\epsfbox[0 0 520 520]{nikolai_e.ps}  
%\epsfxsize=.55\textwidth\epsfbox[0 0 520 520]{nikolai_pt.ps} }
\hspace{-1.cm}\epsfxsize=.55\textwidth\epsfbox[0 0 520 520]{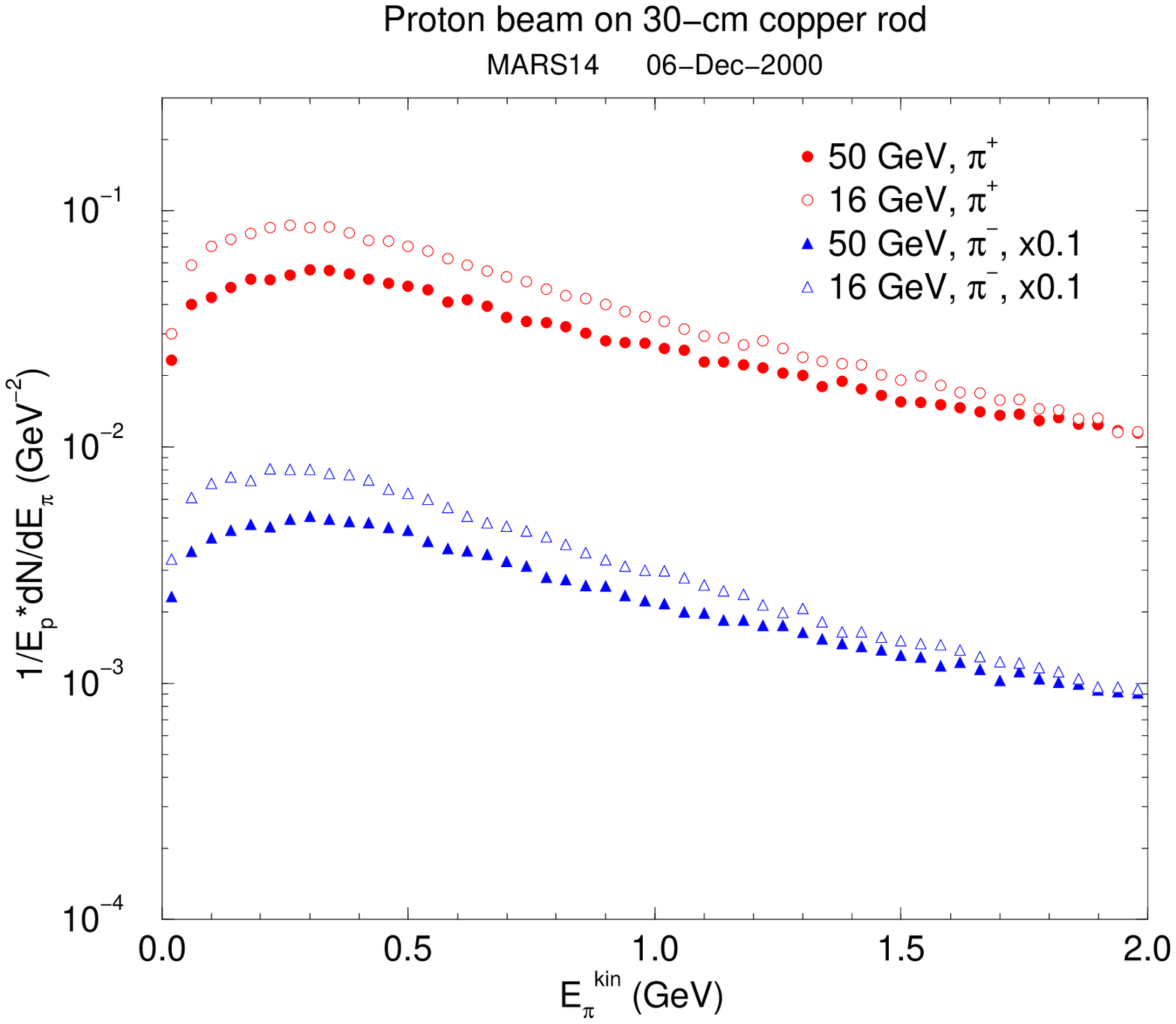}  
\epsfxsize=.55\textwidth\epsfbox[0 0 520 520]{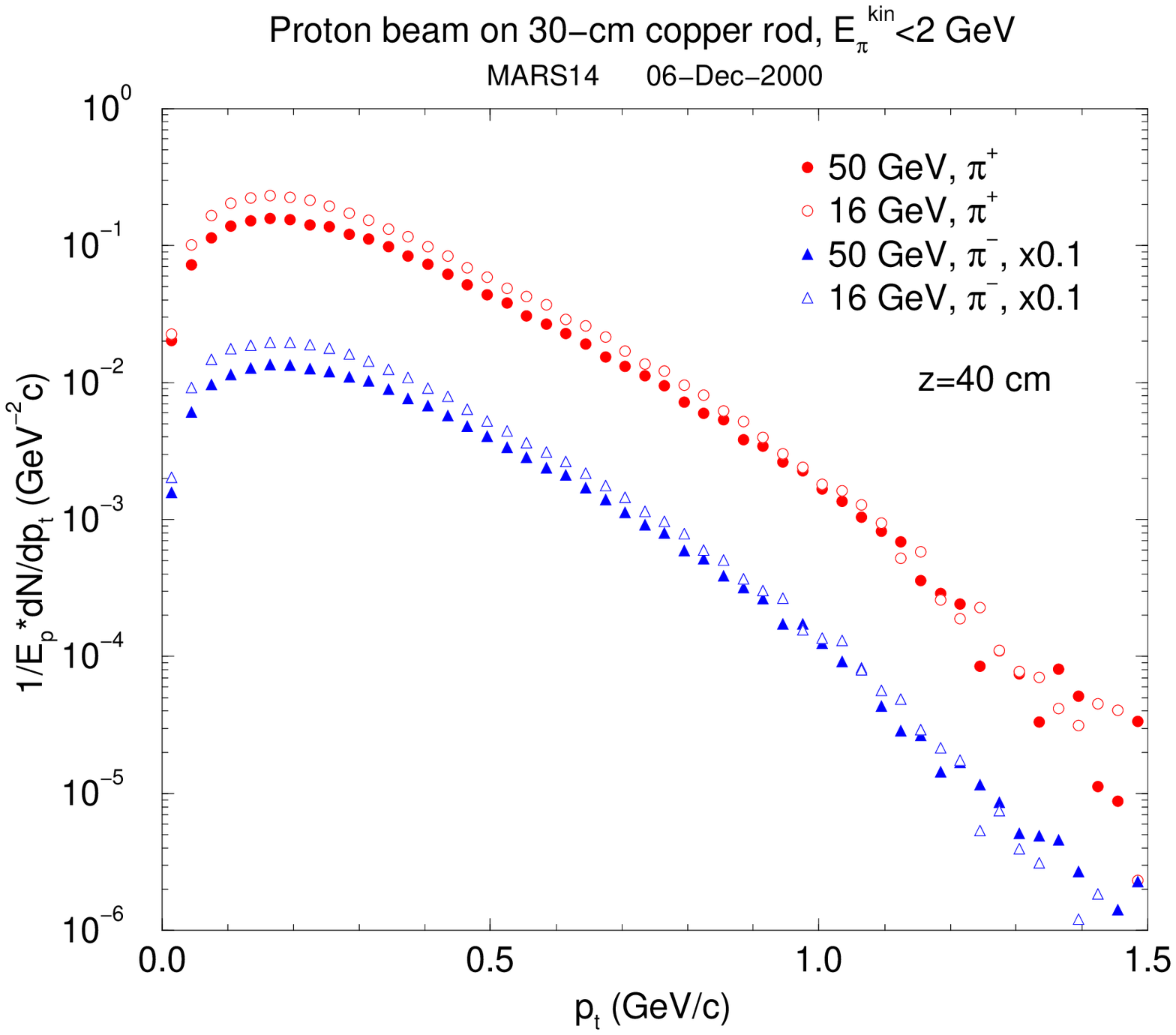} }
\vspace{-0.9cm}
\caption{Charged pion production for 16~GeV and 50~GeV protons 
incident on a 30~cm copper target. The $\pi^+$ and $\pi^-$ kinetic--energy 
(left) and transverse momentum (right) spectra are shown 
for forward going 
particles within a cone of angle 1~radian.
To facilitate a comparison at equal beam power the rates have been 
divided by the incident proton energy. For 
clarity, the $\pi^-$ spectra have been scaled down by an additional 
factor of 10.
}  
\label{fig:16vs50} 
\end{figure}

\subsection{$\sin^2 2\theta_{13}$ reach}

The JHF study group concluded that, 
if no $\nu_e$ appearance signal was observed after 5~years of running, 
taking the central SuperK value for $\delta m^2_{32}$, 
they would be able to exclude $\nu_\mu \to \nu_e$ 
oscillation amplitudes greater than 0.01 at 90\% C.L., which is about an order 
of magnitude better than the present limit 
(Fig.~\ref{fig:jhfloi} right panel). This level of sensitivity corresponds 
to a $\sin^2 2\theta_{13}$ reach of $\sim 0.03$. With a larger dataset 
(20 times SuperK) the reach is improved by about a factor of 4.  

It is interesting to see where the JHF scenario lies in the 
$(f_B, D)$--plane shown in Fig.~\ref{fig:jhfcontours}. 
If we take $f_B = 0.03$ and $D = 77$~kt--years (5~years of 
data taking in the superK detector with a signal efficiency of 0.68), 
we see that the JHF $\to$ SuperK scenario lies close to the 
$\sin^2 2\theta_{13} = 0.03$ contour. Thus the calculations of 
Ref.~\cite{geersb} are in agreement with the JHF study group results. 
Note that 
the JHF scenario already lies in the background systematics 
dominated (vertical contour) region of the plane. 
Upgrading the detector mass by a large factor only 
results in a modest improvement in the $\sin^2 2\theta_{13}$ reach.
With $D = 2600$~kt--years and $f_B = 0.02$, 
the reach has improved to $\sin^2 2\theta_{13} \sim 0.01$ at 
the 0.77~MW JHF. It is unclear whether an upgraded 4~MW JHF would 
further improve the reach, which is very sensitive to $\sigma_{f_B}/f_B$ 
in the systematics dominated region of the ($f_B, D$)--plane. 
A liquid argon detector, with $f_B = 0.003$, $\sigma_{f_B}/f_B = 0.05$, 
and $D = 170$~kt--years, would 
obtain a reach of $\sim 0.01$ at the 0.77~MW JHF, and 
$\sim 0.003$ at an upgraded 4~MW JHF.

To a good approximation we would expect the JHF~study results to  
apply also to a 1~GeV neutrino beam generated at Fermilab using a 16~GeV 
$\sim 1$~MW proton driver. Charged pion production spectra for 16~GeV and 
50~GeV protons are compared in Fig.~\ref{fig:16vs50}, with the spectra 
normalized by dividing by the proton beam energies. Hence the pion event 
rates are shown at equal beam powers. The shapes of the 16~GeV and 
50~GeV kinetic energy-- and transverse--momentum--distributions are 
similar. At equal beam power, the sub-GeV pion rates obtained with 
16~GeV protons are approaching a factor of 
two higher than the 50~GeV rates. Above 1~GeV the rates obtained with 16~GeV 
protons are similar to those obtained with 50~GeV protons if the beam powers 
are similar. 
Hence we would expect the 1~GeV neutrino 
beam fluxes at the JHF to be similar to the fluxes at 
an $\sim 1$~MW 16~GeV machine.

\subsection{Searching for CP-violation} 

     If $\sin^2 2\theta_{13}$ lies within an order of magnitude of 
the present experimental limits, we would expect a $\nu_\mu \to \nu_e$ 
signal to be established in either the next generation of accelerator 
based neutrino experiments, or at a future superbeam experiment.
In this case, if the solar neutrino deficit is correctly described by 
the LMA MSW solution there is the tantalizing possibility of observing 
CP--violation in the lepton sector, and measuring the CP--violating 
amplitude. In a medium baseline experiment 
the CP--violating signature (an asymmetry between the  
$\nu_\mu\to\nu_e$ and $\bar\nu_\mu \to\bar\nu_e$ oscillation probabilities) 
is not complicated by matter effects, which are very small.

\begin{figure}
%\hspace{-.5in}
\centerline{
\epsfxsize=0.9\textwidth 
%\epsffile[0 0 510 320]{new_fig8ab.eps} }
\epsffile[0 0 510 320]{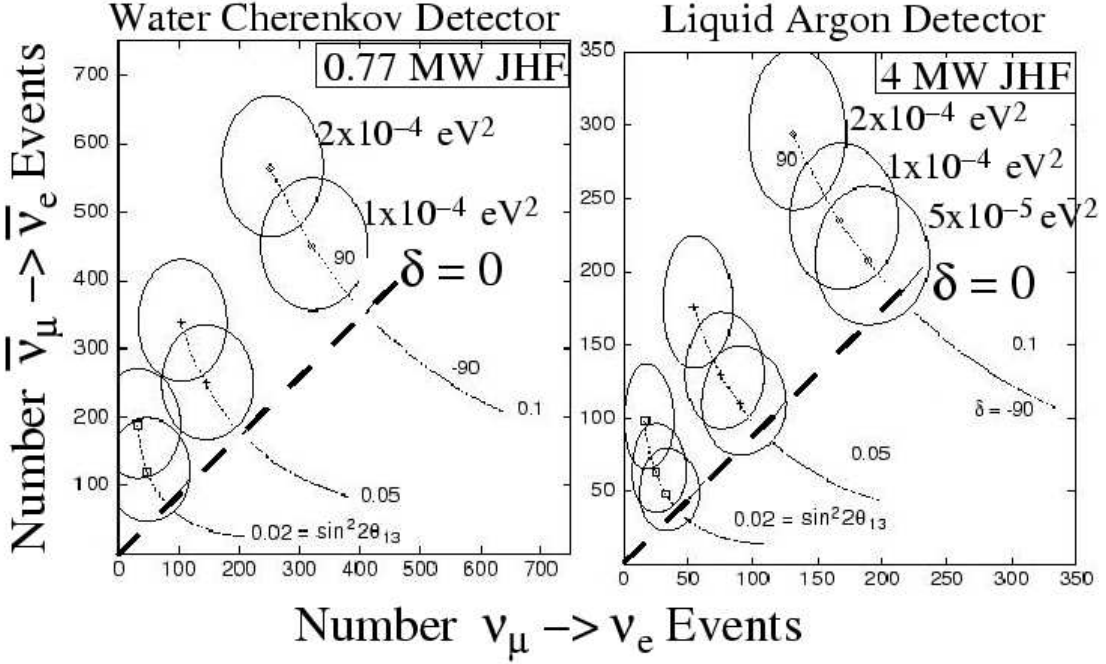} }
%\vspace{-10.0cm}
\caption{
Three--sigma error ellipses in the 
$\left(N_+, N_-\right)$--plane, 
where $N_-$ is the number of $\nu_\mu \to \nu_e$ signal events 
and $N_+$ is the number of $\bar\nu_\mu \to \bar\nu_e$ signal events,
shown for 1~GeV neutrino beams with $L = 295$~km at the 0.77~MW JHF 
using a 220~kt water cerenkov detector (left panel) with 
$f_B = 0.02$, and at the 
4~MW upgraded JHF with a 30~kt liquid argon detector (right panel) 
with $f_B = 0.004$. 
The 3 families of ellipses correspond to $\sin^22\theta_{13} =
0.02$, $0.05$, and $0.1$, as labelled. 
The solid (dashed) [dotted] curves correspond
to $\delta = 0^\circ$ ($90^\circ$) [$-90^\circ$] with $\delta m^2_{21}$
varying from $2\times10^{-5}$~eV$^2$ to $2\times10^{-4}$~eV$^2$.
The error ellipses are shown on each curve for 
$\delta m^2 = 5\times10^{-5}$, $10^{-4}$ and
$2\times10^{-4}$~eV$^2$. The curves assume 3 years of neutrino running 
followed by 6~years of antineutrino running~\cite{geersb}.
} 
\label{fig:jhfcp}
\end{figure}

Consider the sensitivity to CP--violation at the JHF with $L = 295$~km. 
Both the background levels 
and the associated systematic uncertainty are expected to be worse  
for antineutrino beams than for neutrino beams. 
In the following, for simplicity 
we will consider backgrounds and systematics to be the same for $\nu$ 
and $\bar\nu$ beams, and take $f_B = 0.02$ and $\sigma_{f_B}/f_B = 0.1$.
Figure~\ref{fig:jhfcp} shows the expected sensitivity to maximal 
CP--violation ($\delta = 90^\circ$) after 3 years of neutrino running 
to measure the number of $\nu_\mu \to \nu_e$ events, followed by 
6 years of antineutrino running to measure the number of 
$\bar\nu_\mu \to \bar\nu_e$ events. In the absence of CP--violation 
we would expect the two signal samples to have the same number of 
events on average since the factor of two difference in neutrino and 
antineutrino cross--sections is compensated by the difference in the 
running times. Hence the broken curves at $45^\circ$ in the figure correspond 
to the CP--conserving case. 
The figure shows $3\sigma$ error ellipses for a water cerenkov detector
with a fiducial mass of 220~kt at the 0.77~MW JHF (left panel), 
and a liquid argon TPC with a fiducial mass of 30~kt at a 4~MW upgraded JHF 
(right panel). The error ellipses are shown for three different sub--leading 
scales $\delta m^2_{21} = 2 \times 10^{-4}$, $1 \times 10^{-4}$, and 
$5 \times 10^{-5}$~eV$^2$. In each panel the three families of ellipses 
correspond to three values of $\sin^2 2\theta_{13}$. Note that parameter 
values with ellipses 
that lie entirely above the $\delta = 0$ line would result in a $3\sigma$ 
observation of maximal CP--violation. 
We see that for the water cerenkov scenario, if $\sin^2 2\theta_{13} = 0.1$, 
marginally below the currently excluded region, then maximal CP--violation 
would be observed provided $\delta m^2_{21}$ is not significantly below  
$1 \times 10^{-4}$~eV$^2$, which is at the upper end of the preferred 
solar neutrino deficit LMA region. The sensitivity is only marginally better 
in the liquid argon scenario. With decreasing $\sin^2 2\theta_{13}$ the 
sensitivity slowly decreases, with $1 \times 10^{-4}$~eV$^2$ being the 
limiting $\delta m^2_{21}$ for $\sin^2 2\theta_{13} \sim 0.02$. 
Hence, for a small region of presently favored MSW LMA parameter space, 
maximal CP Violation could be seen at 
$3\sigma$ at a 1~GeV medium baseline superbeam. 
This small exciting piece of parameter space can be described approximately 
by: 
\begin{equation}
\sin^2 2\theta_{13} > 0.02 \; , \; \; \; \; \; \; 
\sin\delta \times \delta m^2_{21} > 7 \times 10^{-5} \; {\rm eV}^2
\end{equation}

\section{Conclusions} 

Neutrino superbeams that exploit MW--scale proton drivers, 
together with detectors that are an order of magnitude larger 
than those presently foreseen, offer the 
prospect of improving the sensitivity to $\nu_\mu \to \nu_e$ 
oscillations by an order of magnitude beyond the next generation of 
experiments. Superbeams would therefore provide a useful tool en route 
to a neutrino factory. Our main conclusions are:
\begin{description}
\item{(i)} We believe that the initial $\nu_e$ contamination in the beam 
might be reduced to $\sim 0.2$\%, although we note that the contributions 
from $K_L$ decay will make this goal difficult to acheive for 
multi--GeV beams. 
\item{(ii)} The dominant 
$\nu_\mu \to \nu_e$ backgrounds will arise from (a) $\pi^0$ production in 
NC events, where the $\pi^0$ subsequently fakes an electron signature, 
and (b) $\nu_\tau$ CC interactions (if the beam energy is above $\sim 5$~GeV).
\item{iii)} 
Of the detector technologies we have considered, only the liquid argon 
detector offers the possibility of reducing the background fraction 
$f_B$ significantly below 0.01. A multi--GeV long baseline superbeam 
experiment with a liquid argon (water cerenkov) 
detector would be able to observe a 
$\nu_\mu \to \nu_e$ signal with a significance of at least $3\sigma$ 
above the background provided $\sin^2 2\theta_{13} \gsim 0.002-0.003$ (0.003). 
If the 
baseline $L \gsim 3000$~km, the same experiment would also be able to 
determine the sign of $\delta m^2_{32}$ provided 
$\sin^2 2\theta_{13} \gsim 0.01$. However, it seems unlikely that an 
unambiguous signal for CP--violation could be established with a 
multi--GeV superbeam.
\item{(iv)}  A 1~GeV medium baseline superbeam
experiment with a liquid argon detector would be able to observe a 
$\nu_\mu \to \nu_e$ signal with a significance of at least $3\sigma$ 
above the background provided $\sin^2 2\theta_{13} \gsim 0.003$.
The experiment would not be able to determine the sign of $\delta m^2_{32}$, 
but if the LMA MSW solution correctly describes the solar neutrino deficit, 
there is a small region of allowed parameter space for which CP--violation 
in the lepton sector might be established.
\end{description}

\begin{figure}
\centerline{
\epsfxsize=0.9\textwidth\epsfbox{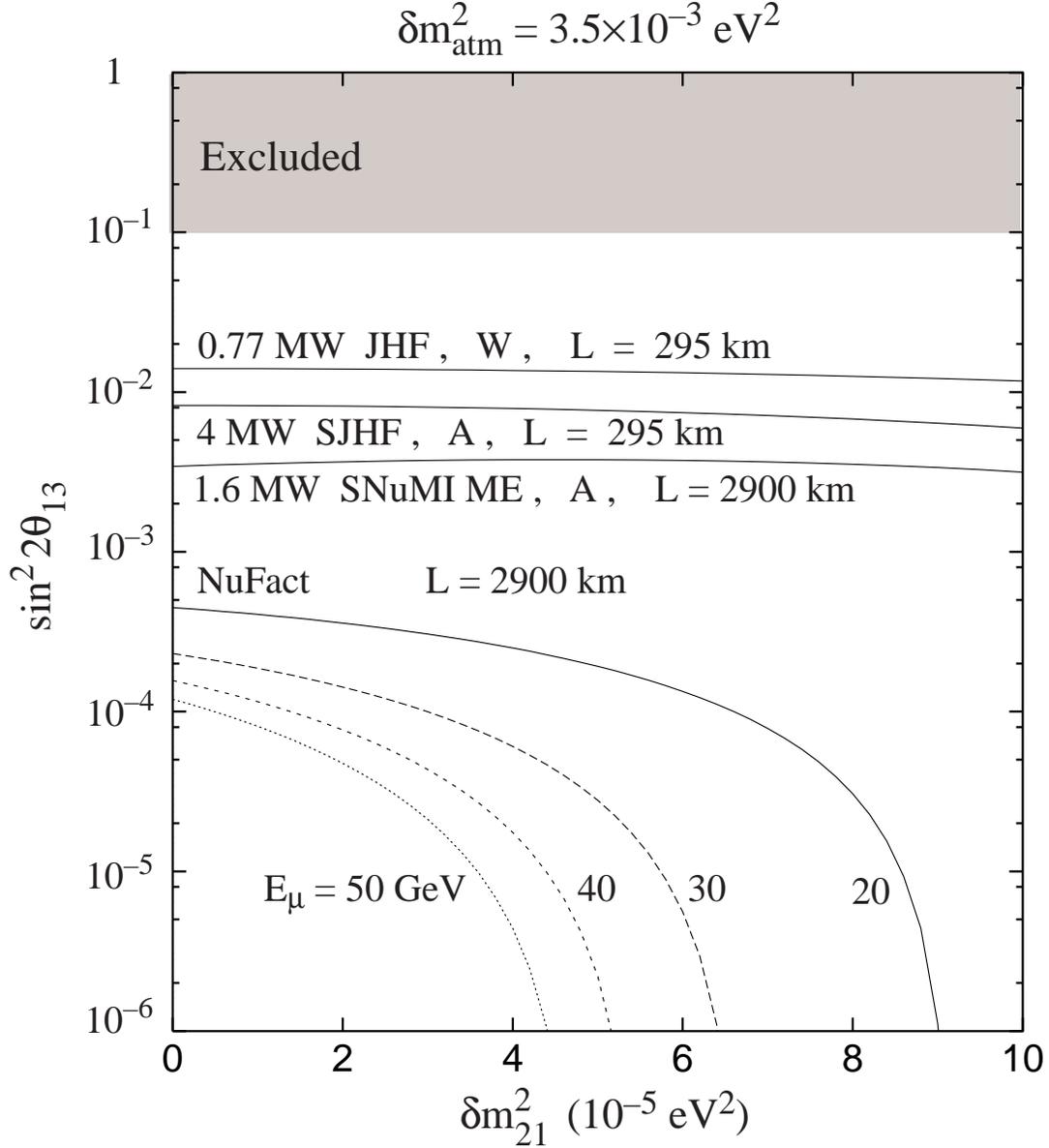} }
\caption{Summary of the $3\sigma$ level sensitivities for the 
observation of $\nu_\mu \to \nu_e$ at various MW--scale superbeams 
(as indicated) with liquid argon ``A'' and water cerenkov ``W'' detector 
parameters, and the observation of $\nu_e \to \nu_\mu$ in a 50~kt detector 
at 20, 30, 40, and 50~GeV neutrino factories delivering $2 \times 10^{20}$ 
muon decays in 
the beam forming straight section. The limiting $3\sigma$ contours are 
shown in the $\delta m^2_{21}, \sin^2 2\theta_{13}$--plane. All curves 
correspond to 3~years of running. The grey shaded 
area is already excluded by current experiments.
}
\label{fig:reach} 
\end{figure}

We compare the superbeam $\nu_\mu \to \nu_e$ reach with the 
corresponding neutrino factory $\nu_e \to \nu_\mu$ reach in 
Fig.~\ref{fig:reach}, which shows the $3\sigma$ sensitivity contours in 
the $(\delta m^2_{21}, \sin^2 2\theta_{13})$--plane. The superbeam 
$\sin^2 2\theta_{13}$ reach of a few $\times 10^{-3}$ is almost independent 
of the sub--leading scale $\delta m^2_{21}$. However, since the neutrino 
factory probes oscillation amplitudes $O(10^{-4})$ the sub--leading effects 
cannot be ignored, and a signal would be observed at a neutrino factory 
over a significant range 
of $\delta m^2_{21}$ even if $\sin^2 2\theta_{13} = 0$.

Finally, both the possibility of exploiting sub--GeV superbeams 
(not considered in our present study), and the optimum detector 
designs for GeV and multi--GeV experiments, deserve 
further consideration. 

\bigskip
\bigskip
{\bf Acknowledegments}
\bigskip

We would like to acknowledge support for this study from the 
U.S. Department of Energy, the Illinois Dept. of
Commerce and Community Affairs, the Illinois State Board of Higher Education,
and the National Science Foundation.

\clearpage

\section{Appendix 1: Neutrino Masses and Mixing}

In this appendix we briefly review the theoretical framework used 
to describe neutrino oscillations.

\subsection{Neutrino mass}

In the standard SU(3) $\times$ SU(2)$_L \times$ U(1)$_Y$ model 
(SM) neutrinos occur in SU(2)$_L$ doublets with $Y=-1$:
\beq 
{\cal L}_{L \ell} = \left ( \begin{array}{c}
                     \nu_\ell \\
                     \ell \end{array} \right )
 \ , \quad  \ell=e, \ \mu, \ \tau 
\eeq
There are no electroweak-singlet neutrinos (often called right-handed
neutrinos)  $\chi_{R,j}$, $j=1,...,n_s$.  Equivalently, these could be 
written as $\overline{\chi^c}_{L,j}$.  There are three types of 
possible Lorentz-invariant bilinear operator products that can be formed from 
two Weyl fermions $\psi_L$ and $\chi_R$: 

\begin{itemize} 

\item 

Dirac: \ $m_D \bar \psi_L \chi_R + h.c.$ \ This connects opposite-chirality
fields and conserves fermion number. 

\item 

Left-handed Majorana: \ $m_L \psi_L^T C \psi_L + h.c.$ where
$C=i\gamma_2\gamma_0$ is the charge conjugation matrix.

\item 

Right-handed Majorana: \ $M_R  \chi_R^T C \chi_R + h.c.$

\end{itemize}
The Majorana mass terms connect fermion fields of the same chirality and
violate fermion number (by two units).  
Using the anticommutativity of fermion fields and the property $C^T = -C$, it
follows that a Majorana mass matrix appearing as 
\beq
\psi_i^T C (M_{maj})_{ij} \psi_j
\eeq
is symmetric in flavor indices: 
\beq
M_{maj}^T = M_{maj} 
\eeq
Thus, in the SM, there is 
no Dirac neutrino mass term because (i) it is forbidden as a bare
mass term by the gauge invariance, (ii) it cannot occur, as do the quark and
charged lepton mass terms, via spontaneous symmetry breaking (SSB) of the
electroweak (EW) symmetry starting from a Yukawa term because there are no
EW-singlet neutrinos $\chi_{R,j}$.  There is also 
no left-handed Majorana mass term because (i) it is forbidden as a bare mass 
term and (ii) it would
require a $I=1$, $Y=2$ Higgs field, but the SM has no such Higgs field.
Finally, there is 
no right-handed Majorana mass term because there is no
$\chi_{R,j}$.  The same holds for the minimal supersymmetric standard model 
(MSSM) and the minimal SU(5) grand unified theory (GUT), both for the original
and supersymmetric versions. 

However, it is easy to add electroweak-singlet neutrinos $\chi_R$ to the SM,
MSSM, or SU(5) GUT; these are gauge-singlets under the SM gauge group and
SU(5), respectively.  Denote these theories as the extended SM, etc. This gives
rise to both Dirac and Majorana mass terms, the former via Yukawa terms and the
latter as bare mass terms.  

In the extended SM, MSSM, or SU(5) GUT, one could consider the addition of the
$\chi_R$ fields as {\it ad hoc}.  However, a more complete grand unification is
achieved with the (SUSY) SO(10) GUT, since all of the fermions of a
given generation fit into a single representation
 of SO(10), namely, the 16-dimensional spinor
representation $\psi_L$.  In this theory the states $\chi_R$ are not {\it
ad hoc} additions, but are guaranteed to exist.  
In terms of SU(5) representations
(recall, SO(10) $\supset$ SU(5) $\times$ U(1))
\beq
16_L = 10_L + \bar 5_L + 1_L
\eeq
so for each generation, in addition to the usual 15 Weyl fermions comprising
the 10$_L$ and $5_R$, (equivalently $\bar 5_L$) of SU(5), there is also an
SU(5)-singlet, $\chi^c_L$ (equivalently, $\chi_R$). 
So in SO(10) GUT, 
electroweak-singlet neutrinos are guaranteed to occur, with number equal to the
number of SM generations, inferred to be $n_s=3$.  Furthermore, 
the generic scale for the coefficients in $M_R$ is expected to be the GUT 
scale, $M_{GUT} \sim 10^{16}$ GeV. 

There is an important mechanism, which originally arose in the context of
GUT's, but is more general, that naturally predicts light neutrinos.  This is
the seesaw mechanism~\cite{seesaw}.  
The basic point is that because the Majorana mass term 
is an electroweak singlet, the associated Majorana 
mass matrix $M_R$ should not be related to the electroweak mass scale $v$, and 
from a top-down point of view, it should be much larger than this scale. Denote
this generically as $m_R$. 
This has the very important consequence that when we diagonalize the joint
Dirac-Majorana mass matrix, the eigenvalues (masses) will be comprised of
two different sets: $n_s$ heavy masses, of order $m_R$, and 3 light
masses.  
The largeness of $m_R$ then naturally explains the smallness of the masses (or,
most conservatively, upper bounds on masses) of the known neutrinos.  
This appealing mechanism also applies in the
physical case of three generations and for $n_s \ge 2$. 

At a phenomenological level, without further theoretical assumptions,
there is a large range of values for the light $m_\nu$, since (1) the
actual scale of $m_R$ is theory-dependent, and (2) it is, {\it a priori}, not
clear what to take for $m_D$ since the known (Dirac) masses range over 5 orders
of magnitude, from $m_e, m_u \sim $ MeV to $m_t = 174$ GeV, and this
uncertainty gets squared.  However, in the SO(10) GUT scheme, where one can
plausibly use $m_D \sim m_t$ for the third-generation neutrino, and 
$m_R \sim M_{GUT} \sim 10^{16}$ GeV for the scale of masses in the right-handed
Majorana mass matrix, one has 
\beq
m(\nu_3) \sim \frac{m_t^2}{m_R} \sim 10^{-3} \ {\rm eV}
\eeq
which is close to the value $m(\nu_3) = 0.06$ eV obtained from 
$\delta m^2_{atm}$ if one assumes a hierarchical neutrino mass spectrum with
$m(\nu_3) >> m(\nu_2)$.  Thus, the seesaw mechanism not only provides an
appealing qualitative explanation of why neutrino masses are much smaller than
the masses of the other known fermions, but also, with plausible assumptions,
predicts a value for $m(\nu_3)$ comparable to suggestions from current
atmospheric neutrino data.  

\subsection{Neutrino mixing and oscillations}

The unitary transformation
relating the mass eigenstates to the weak eigenstates is as follows,  
\beq
\nu_{\ell_a} = \sum_{i=1}^3 U_{a i} \nu_i \ , \quad \ell_1=e, \ \ell_2=\mu, \ 
\ell_3=\tau
\eeq
i.e., 
\beq
\left ( \begin{array}{c} \nu_e \\ \nu_\mu \\ \nu_\tau \end{array} \right ) 
=  \left( \begin{array}{ccc}
              U_{e1} & U_{e2} & U_{e3}  \\
              U_{\mu 1} & U_{\mu 2} & U_{\mu 3} \\
              U_{\tau 1} & U_{\tau 2} & U_{\tau 3} \end{array} \right )
\left( \begin{array}{c} \nu_1 \\
                        \nu_2 \\
                         \nu_3 \end{array} \right ) 
\eeq

One possible representation of this $3 \times 3$ unitary matrix is 
\beq
U=
\pmatrix{c_{12} c_{13} & s_{12}c_{13} & s_{13} e^{-i\delta} \cr
-s_{12}c_{23}-c_{12}s_{23}s_{13}e^{i\delta}
& c_{12}c_{23}-s_{12}s_{23}s_{13}e^{i\delta} & s_{23}c_{13} \cr
s_{12}s_{23}-c_{12}c_{23}s_{13}e^{i\delta}
&-c_{12}s_{23}-s_{12}c_{23}s_{13}e^{i\delta} & c_{23}c_{13}}
\eeq
where 
%$R_{ij}$ is the rotation matrix in the $ij$ subspace,
$c_{ij}=\cos\theta_{ij}$, $s_{ij}=\sin\theta_{ij}$.  
Thus, in this framework, the neutrino mixing depends on the four angles
$\theta_{12}$, $\theta_{13}$, $\theta_{23}$, and $\delta$, 
and on two independent
differences of squared masses, $\delta m^2_{atm.}$, which is 
$\delta m^2_{32} = m(\nu_3)^2-m(\nu_2)^2$ in the favored fit, and 
$\delta m^2_{sol}$, which may be taken to be $\delta m^2_{21}=m(\nu_2)^2-
m(\nu_1)^2$.  Note that these quantities involve both magnitude and sign;
although in a two-species neutrino oscillation in vacuum the sign does not
enter, in the three species oscillations relevant here, and including both
matter effects and CP violation, the signs of the $\delta m^2$ quantities do
enter and can, in principle, be measured.

In the 1980's, most theorists thought that lepton mixing would be hierarchical,
i.e. the lepton mixing matrix $U$ would differ from the identity by small
entries, and these would be smaller as one moved further from the diagonal, as
is established to be the case with quark mixing.  This was, indeed, a large
part of the appeal of the MSW mechanism: it could produce large mixing with
small vacuum mixing angles.  However, the results from the SuperK measurements
of atmospheric neutrinos have forced a revision in this conventional picture,
providing strong evidence for essentially maximal mixing,
$\sin^2 2\theta_{23}=1$.  A challenge to model-builders has thus been to get
maximal $\sin^2 2\theta_{23}$.  More recently, the SuperK solar neutrino data
favors large $\sin^2 2\theta_{12}$.  Bimaximal mixing schemes take
$\theta_{23}=\theta_{12}=\pi/4$ and $\theta_{13} << 1$ \cite{bimax}.  There are
no compelling theoretical suggestions concerning the magnitude of
$\theta_{13}$, and one of the important physics goals for neutrino oscillation
experiments with conventional beams is to try to measure this angle. 

For our later discussion it will be useful to record the formulas for the
various relevant neutrino oscillation transitions.  In the absence of any
matter effect, the probability that a (relativistic) weak neutrino eigenstate
$\nu_a$ becomes $\nu_b$ after propagating a distance $L$ is
\beqs
P(\nu_a \to \nu_b) &=& \delta_{ab} - 4 \sum_{i>j=1}^3
Re(K_{ab,ij}) \sin^2  \Bigl ( \frac{\delta m_{ij}^2 L}{4E} \Bigr ) +
\nonumber\\&+& 4 \sum_{i>j=1}^3 Im(K_{ab,ij})
 \sin \Bigl ( \frac{\delta m_{ij}^2 L}{4E} \Bigr )
\cos \Bigl ( \frac{\delta m_{ij}^2 L}{4E} \Bigr )
\label{pab}
\eeqs
where
\beq
K_{ab,ij} = U_{ai}U^*_{bi}U^*_{aj} U_{bj}
\label{k}
\eeq
and
\beq
\delta m_{ij}^2 = m(\nu_i)^2-m(\nu_j)^2
\label{delta}
\eeq
Recall that in vacuum, CPT invariance implies
$P(\bar\nu_b \to \bar\nu_a)=P(\nu_a \to \nu_b)$ and hence, for $b=a$,
$P(\bar\nu_a \to \bar\nu_a) = P(\nu_a \to \nu_a)$.  For the
CP-transformed reaction $\bar\nu_a \to \bar\nu_b$ and the T-reversed
reaction $\nu_b \to \nu_a$, the transition probabilities are given by the
right-hand side of (\ref{pab}) with the sign of the imaginary term reversed.
In the following we will assume CPT invariance, 
so that CP violation is equivalent to T violation. 

The solar and atmospheric neutrino data indicate that 
\beq 
\delta m^2_{21}
= \delta m^2_{sol} \ll \delta m^2_{31} \approx \delta m^2_{32}=\delta m^2_{atm}
\label{hierarchy}
\eeq
In this case, CP (T) violation effects are very small, so that in
vacuum
\beq
P(\bar\nu_a \to \bar\nu_b) \simeq P(\nu_a \to \nu_b)
\label{pcp}
\eeq
\beq
P(\nu_b \to \nu_a) \simeq P(\nu_a \to \nu_b)
\label{pt}
\eeq
In the absence of T violation, the second equality (\ref{pt}) would still hold
in matter, but even in the absence of CP violation, the first equality
(\ref{pcp}) would not hold.  With the hierarchy (\ref{hierarchy}), the
expressions for the specific oscillation transitions are
\beqs
P(\nu_\mu \to \nu_\tau) & = & 4|U_{33}|^2|U_{23}|^2
\sin^2 \Bigl ( \frac{\delta m^2_{atm}L}{4E} \Bigr ) \cr\cr
& = & \sin^2 2\theta_{23} \cos^4 \theta_{13}
\sin^2 \Bigl (\frac{\delta m^2_{atm}L}{4E} \Bigr )
\label{pnumunutau}
\eeqs
\beqs
P(\nu_\mu \to \nu_e) & = & 4|U_{13}|^2 |U_{23}|^2
\sin^2 \Bigl ( \frac{\delta m^2_{atm}L}{4E} \Bigr ) \cr\cr
& = & \sin^2 2\theta_{13}\sin^2 \theta_{23}
\sin^2 \Bigl (\frac{\delta m^2_{atm}L}{4E} \Bigr )
\label{pnuenumu}
\eeqs
With units inserted, one has the general relation 
\beq
\sin^2 \Bigl (\frac{\delta m^2L}{4E} \Bigr ) = 
\sin^2 \Bigl (\frac{1.27 (\delta m^2/{\rm eV}^2)(L/{\rm km})}
{(E/{\rm GeV})} \Bigr ) 
\label{oscrel}
\eeq
This makes it clear what the approximate sensitivity of an experiment with a
given pathlength is to a neutrino oscillation channel involving a given $\delta
m^2$, for a beam with an energy $E$. 

There can be significant corrections to the one-$\delta m^2$ oscillation 
formulas if $\delta m^2_{sol}$ is at the upper end of the LMA range, 
$\delta m^2_{sol} \sim 10^{-4}$ eV$^2$, 
if $\sin^2 2\theta_{13}$ is sufficiently 
small.  In this case, keeping dominant terms and neglecting possible small CP 
violating terms, eq. (\ref{pnuenumu}) becomes 
\beqs
P(\nu_\mu \to \nu_e) & = & \sin^2 2\theta_{13} \sin^2 \theta_{23}
\sin^2 \Bigl (\frac{\delta m^2_{atm}L}{4E} \Bigr ) \cr\cr
& + & \sin^2 2\theta_{12} \cos^2 \theta_{13} \cos^2 \theta_{23}
\sin^2 \Bigl (\frac{\delta m^2_{sol}L}{4E} \Bigr ) 
\label{pnuenumuc}
\eeqs
Let us denote the two terms as $T_1$ and $T_2$.  As an illustrative example,
let us consider a pathlength $L$ sufficiently short that matter effects are not
too important.  Assume $\sin^2 2\theta_{13} =0.01$ and the upper end 
of the LMA solution, with $\sin^2 2\theta_{12} =0.8$ and $\delta m^2_{sol}
= 10^{-4}$ eV$^2$.  Then for $L=730$ km, $T_2 = 0.1 T_1$.  For these values,
$\sin^2(\delta m^2_{atm}L/(4E))=0.78$ while 
$\sin^2(\delta m^2_{sol}L/(4E))=0.95 \times 10^{-3}$, so that the pathlength is
causing a strong suppression of the subdominant oscillation due to $\delta
m^2_{sol}$.  
%For a calculation with $L=2900$ km, taking into account matter
%effects, see, e.g., Fig. 5 of \cite{lb}.  
For sufficiently large $L$ and small
$\sin^2 2\theta_{13}$, the $\delta m^2_{sol}$ oscillation can be a significant
contribution to $\nu_\mu \to \nu_e$.  However, we note that making $L$ greater
would mean that one would also have to make $E$ greater to keep an acceptable
event rate with a given detector, and this would tend to increase backgrounds
to the $\nu_\mu \to \nu_e$ signal.  We also note that if 
KamLAND~\cite{kamland} achieves its
projected sensitivity, it will have tested the LMA solution by $\sim$ 2005. 

In neutrino oscillation searches using reactor antineutrinos,
i.e. tests of $\bar\nu_e \to \bar\nu_e$, the two-species mixing hypothesis used
to fit the data is
\beq
P(\nu_e \to \nu_e) = 1 - \sin^2 2\theta_{reactor}
\sin^2 \Bigl (\frac{\delta m^2_{reactor}L}{4E} \Bigr )
\label{preactor}
\eeq
where $\delta m^2_{reactor}$ is the squared mass difference relevant for
$\bar\nu_e \to \bar\nu_x$.  In particular, in the upper range of values of
$\delta m^2_{atm}$, since the transitions $\bar\nu_e \to \bar\nu_\mu$ and
$\bar\nu_e \to \bar\nu_\tau$ contribute to $\bar\nu_e$ disappearance, one has
\beq
P(\nu_e \to \nu_e) = 1 - \sin^2 2\theta_{13} \sin^2 \Bigl
(\frac{\delta m^2_{atm}L}{4E} \Bigr )
\label{preactoratm}
\eeq
i.e., $\theta_{reactor}=\theta_{13}$, and the Chooz reactor experiment yields
the bound~\cite{chooz}
\beq
\sin^2 2\theta_{13} < 0.10
\label{chooz}
\eeq
which is also consistent with conclusions from the SuperK data analysis
\cite{superk}.

Further, the quantity ``$\sin^2 2\theta_{atm}$'' often used to fit
the data on atmospheric neutrinos with a simplified two-species mixing
hypothesis, is, in the three-generation case,
\beq
\sin^2 2\theta_{atm} \equiv \sin^2 2\theta_{23} \cos^4 \theta_{13}
\label{thetaatm}
\eeq
The SuperK data implies that (up to redefinitions of quadrants, etc.)
\beq
\theta_{23} \simeq \frac{\pi}{4}
\label{theta23}
\eeq
and $\sin^2 2\theta_{13} << 1$.  Thus, to
good accuracy, $\theta_{atm} = \theta_{23}$.

The types of neutrino oscillations that can be searched for with a conventional
neutrino beam include: 

\begin{itemize}

\item

$\nu_\mu \to \nu_\mu$ (disappearance)

\item

$\nu_\mu \to \nu_e$, $\nu_e \to e^-$ (appearance)

\item 

$\nu_\mu \to \nu_\tau$, $\nu_\tau \to \tau^-$; $\tau^- \to (e^-, \mu^-)...$
 (appearance)

\end{itemize}
Searches for the conjugate oscillation channels require $\bar\nu_\mu$ beams. 
Since these
have lower fluxes than $\nu_\mu$ beams (and this difference can be large
with sign-selected $\pi$ beams that are decaying), one can concentrate on
oscillation channels with $\nu_\mu$ beams.  

For neutrino oscillation experiments with pathlengths of order $10^3$ km,
matter effects are significant.  These have been studied in a number of papers,
e.g., \cite{bernpark,petcov,akh,bargergeer,bgrw2,lb}.  The constant density
assumption provides a first approximation; realistic density profiles were
included in the calculations of \cite{bargergeer,bgrw2,lb}.  
In the constant density
approximation, for a simple two-species mixing, one has
\beq
P(\nu_a \to \nu_b) = \sin^2(2\theta_m)\sin^2(\omega L)
\label{pmatter}
\eeq
where
\beq
\sin^2(2\theta_m) = \frac{\sin^2(2\theta)}{\sin^2(2\theta) + \biggl [
\cos(2\theta) - \frac{2\sqrt{2}G_FN_e E}{\delta m^2} \biggr ]^2}
\label{sin2m}
\eeq
\beq
\omega^2 = \biggl [ \frac{\delta m^2}{4E}\cos(2\theta) 
- \frac{G_f}{\sqrt{2}}N_e \biggr ]^2 + 
\biggl [ \frac{\delta m^2}{4E}\sin(2\theta) \biggr ]^2
\label{omega}
\eeq
where $N_e$ is the electron number density of the matter. 

\bigskip

\section{Appendix 2: Detector unit costs}

Estimating the unit costs for each detector type is not
straightforward. We can base our cost estimates on detectors that are
currently under construction, or have recently been proposed. However,
these example detectors have been proposed/costed at different times
using different accounting systems in different currencies with
different levels of external scrutiny.  To attempt to compare
like--with--like we have started from the bare materials and services
(M\&S) costs of the detector itself, which we have corrected to
include salaries (SWF), engineering and R\&D (EDIA) costs. The scaling
factors were determined for a current US-based neutrino detector
(MINOS).  An estimate of overheads and contingencies (35\%) has been
included to reflect the ``fully-loaded'' costs associated with a
US-based detector.  Finally, the resulting unit costs have been
corrected for inflation to correspond to FY01 dollars. Based on the
fully loaded FY01 unit costs, for each detector type the mass of the
detectors that could be built with a budget of \$500M can be
estimated. The costs for a cavern for each detector technology is 
based on the recent UNO estimates for a hard rock cavern
(\$200/m$^3$) using the computed detector masses and the densities of
the various detector media~\cite{uno}. The results are summarized in
Table~\ref{taba2}.

The bare detector costs are based on the following:
\begin{description}

 \item{(i)} The water cherenkov detector estimates are based on those 
documented in the UNO cost estimate~\cite{uno}, and correspond to 237M\$/450~kt
(FY00 dollars), assuming 10\% photomultiplier coverage in the entire
detector. We assume a hard rock site rather than the proposed
WIPP site. The actual UNO proposal is based on 40\% photomultiplier
coverage in the central third of the detector's volume, which is 
optimized for certain proton decay and astrophysical neutrino channels.
These physics topics are the driving force behind the UNO proposal.

 \item{(ii)} The liquid argon unit cost is based on the Icanoe costs of
14.4~MEuro per 1.9~kt module, with \$0.9425 per/Euro (FY99 dollars).
Note the costing presented in this document assumes only cryogenic
modules~\cite{icacost}.

 \item{(iii)} The steel-scintillator unit cost is based on the 
MINOS M\&S unit cost, which is based on the most recent
(2/01) far-detector cost data giving a total of 16.3M\$ (FY98
dollars)~\cite{minostdr,jeff}.  The total far detector (5.4~kt) cost
including R\&D, labor, and institutional overhead costs in then-year
dollars is 25.4M\$.

 \item{(iv)} The mineral oil cerenkov unis cost is based on the 
MiniBooNE M\&S unit cost, which is based on TDR detector
costs (FY00 dollars)~\cite{miniboone,ray}.  

\end{description}

\begin{table}[h]
\caption{Detector cost estimates.} 
\begin{tabular}{lcccc}
&&&&\\
\hline
 &  Water &Mineral Oil& Liquid & Steel/\\
 &Cerenkov&Cerenkov& Argon  &Scintillator \\        
% & (UNO)  &  (BooNE)   &(ICARUS)& (MINOS) \\
\hline
%Mass (kt)                 &  650  & 650 &  1.9   &   5.4   \\
%Bare Cost (M\$)           &  237  & 843 &  13.6  &   16.3  \\
Bare Unloaded Unit Cost (M\$/kt)   &  0.36 & 1.3 &   7.1  &   3.0   \\
Unloaded Unit Cost (M\$/kt)~$^{a)}$   &  0.57 & 1.75&  11.2  &   4.7   \\
FY for estimates          & 2000  & 2000& 1999   & 1998    \\
\hline
Loaded Unit Cost (M\$/kt)~$^{b)}$ & 0.67 & 1.92 & 13.5 & 5.9  \\ 
Mass (kt) per \$500M       & 745  & 261 & 37 & 85            \\
\hline
Medium density (g/cm$^3$) &1.0 & 0.9 & 1.8 & 3.5        \\
Cavern cost (M\$)~$^c)$ & 106  & 41 & 2.9 & 3.4           \\
\hline \\
\end{tabular}
\\\\
a) M\&S + SWF + EDIA \\
b) FY01 costs including  overhead and 35\% contingency \\
c) Note that deep caverns are not necessarily needed.
\label{taba2}
\end{table}


\clearpage

\begin{thebibliography}{99}

%[1]
\bibitem{report}
C.~Albright et al., ``Physics at a Neutrino Factory'', 
report to the Fermilab Directorate, FERMILAB-FN-692, 
April 2000.

%[2]
\bibitem{nufact}
S.~Geer, Phys. Rev. {\bf D57}, 6989 (1998).

%[3]
\bibitem{norbert}N. Holtkamp, D.A. Finley, {\em et al},
FERMILAB-PUB-00-108-E, June 2000, submitted to 
Phys.Rev.ST Accel.Beams.  

%[4]
\bibitem{jhfloi} Y.~Itow {\em et al}, Japanese Hadron Facility Letter 
of Intent, February 2000.  

%[5]
\bibitem{k2k}
K.~Nishikawa et al. (KEK-PS E362 Collab.),
``Proposal for a Long Baseline Neutrino Oscillation Experiment,
using KEK-PS and Super-Kamiokande", 1995, unpublished;
talk presented by M.~Sakuda (K2K collaboration) at the {\it XXXth  
International Conference on High Energy Physics (ICHEP~2000)}, Osaka,
Japan, July 2000.

%[6]
\bibitem{minostdr} MINOS Technical Design Report 
NuMI-L-337 October 1998.  

%[7]
\bibitem{opera}
See the OPERA web page at http://www.cern.ch/opera/

%[8]
\bibitem{icarus}
See the ICARUS/ICANOE web page at http://pcnometh4.cern.ch/

%[9]
\bibitem{lsnd}
C.~Athanassopoulos et al. (LSND Collab.),
Phys. Rev. Lett. {\bf 77}, 3082 (1996); {\bf 81}, 1774 (1998);
%\bibitem{LSND2}
G.~Mills, talk at {\it Neutrino-2000}, XIXth International Conference
on Neutrino Physics and Astrophysics, Sudbury, Canada, June 2000.

%[10]
\bibitem{miniboone}
A.~Bazarko, MiniBooNE Collaboration, talk at {\it Neutrino-2000}., 
S.~Koutsoliotas et al., The MiniBooNE Technical Design Report.

%[11]
\bibitem{fpp} These plots have been produced using 
the fast Monte Carlo simulation 
which is currently being used for the CNGS neutrino fluxes. 
The hadron production model is 
the BMPT parameterization which is thought to be the most complete 
parameterization of the existing data: 
M. Bonesini, A. Marchionni, F. Pietropaolo, T. Tabarelli de Fatis, 
hep-ph/0101163



%[12]
\bibitem{numitdr} NumI Facility Technical Design Report NuMI-346 March 1998.

%[13]
\bibitem{geersb} V. Barger, S. Geer, R. Raja, K. Whisnant, hep-ph/0012017.  

%[14]
\bibitem{orland}
See the ORLaND web page  http://www.phys.subr.edu/orland/

%[15]
\bibitem{richter}{B. Richter, hep-ph/0008222.} 

%[16]
\bibitem{fmcgeant} Reference for GEANT and LEPTO here...

%[17]
\bibitem{cngs}
CNGS conceptual design report, CERN 98--02.

%[18]
\bibitem{icanoetdr} The ICARUS and NOE collaborations, 
ICANOE Proposal, chapters 4 and 5, CERN/SPSC 99-25 (1999). 

%[19]
\bibitem{dcasper} D. Casper, private communication.  

%[20]
\bibitem{wai} L. Wai, S. Wojcicki, B. Patterson, 
NuMI-L-713, hep-ph/0101090, and L. Wai, private communication.

%[21]
\bibitem{maury} H.M. Gallagher and M.C.Goodman, 
``Neutrino Cross Sections'', NuMI-112, 1995.

%[22]
\bibitem{icanoe2} Second addendum to ICANOE proposal, 
CERN/SPSC 99-40 (Nov. 1999)

%[23]
\bibitem{p907} See P907~Proposal,  
http://ppd.fnal.gov/experiments/e907/p907/p907.ps

%[24]
\bibitem{harp} Proposal CERN/PS~214; see the HARP web page 
http://harp.web.cern.ch/harp/

%[25]
\bibitem{nomad} See the NOMAD web page 
http://nomadinfo.cern.ch/

%[26]
\bibitem{cousins} S. Mishra, private communication and NOMAD NIM paper in 
preparation.  

%[27]
\bibitem{uno} C.K.Jung, 
%FEASIBILITY OF A NEXT GENERATION UNDERGROUND WATER CERENKOV DETECTOR: 
(UNO Collab.)  \verb+hep-ex/0005046+., and talk at the NNN00 workshop, 
August~2000. 

%[28]
\bibitem{icacost} F.Arneodo {\em et al.} (ICARUS and NOE Collab.),
``ICANOE: Imaging and calorimetric neutrino oscillation experiment''
CERN/SPSC 99-25, Addendum 1: Preliminary Technical Design \& Cost
Estimates, CERN/SPSC 99-39, both of which 
can be found at \verb+http://pcnometh4.cern.ch+

%[29]
\bibitem{soudancost} Rough cost estimate, M.Goodman, private communication.

%[30]
\bibitem{e734cost} Rough cost estimate, Milind Diwan, private communication. 

%[31]
\bibitem{superk}
Super-Kamiokande Collaboration,
Y.~Fukuda et al., Phys. Lett. {\bf B433}, 9 (1998);
Phys. Lett. {\bf B436}, 33 (1998);
Phys. Rev. Lett. {\bf 81}, 1562 (1998);
Phys. Rev. Lett. {\bf 82}, 2644 (1999).

%[32]
\bibitem{seesaw}{M. Gell-Mann, R. Slansky, and P. Ramond, in {\it Supergravity}
(North-Holland, 1979), p. 315; T. Yanagida, in {\it Proceedings of the Workshop
on Unified Theory and Baryon Number in the Universe} (KEK, Japan, 1979).}

%[33]
\bibitem{bimax}{V. Barger et al., Phys. Lett. {\bf B437}, 107 (1998);
A. Baltz, A. S. Goldhaber, and M. Goldhaber, Phys. Rev. Lett. 
{\bf 81}, 5730 (1998); R. Mohapatra and S. Nussinov, 
Phys. Rev. {\bf D60}, 013002 (1999); G. Altarelli and F. Feruglio, Phys. 
Lett. {\bf B439}, 112 (1998); C. Albright, K. Babu, and S. Barr, 
Phys. Rev. Lett. {\bf 81}, 1167 (1998); C. Albright and S. Barr, 
Phys. Lett. {\bf B461}, 218 (1999).} 

%[34]
\bibitem{kamland}
KamLAND proposal, Stanford--HEP--98--03; A. Piepke, talk at {\it
Neutrino--2000}, XIXth International Conference on Neutrino Physics and
Astrophysics, Sudbury, Canada, June 2000.

%[35]
\bibitem{chooz}{M. Apollonio et al., Phys. Lett. {\bf B420}, 397 (1998);
Phys. Lett. {\bf B466}, 415 (1999).} 

%[36]
\bibitem{bernpark}{R. Bernstein and S. Parke, Phys. Rev. {\bf D44}, 2069 
(1991).}

%[37]
\bibitem{petcov}{ S. Petcov, Phys. Lett. {\bf B434}, 321 (1998).  M. Chiznov,
M. Maris, S. Petcov, hep-ph/9810501; M. Chiznov, S. Petcov, hep-ph/9903424;
M.Chiznov, S.Petcov, Phys. Rev. Lett. 83,1096 (1999).}

%[38]
\bibitem{akh}{E. Akhmedov, A. Dighe, P. Lipari, A. Smirnov, Nucl. Phys. {\bf
B542}, 3 (1999); E. Akhmedov, Nucl.Phys. {\bf B538}, 25 (1999);
hep-ph/0001264.} 

%[39]
\bibitem{bargergeer}{V. Barger, S. Geer, K. Whisnant, Phys.Rev. {\bf D61},
053004 (2000).}

%[40]
\bibitem{bgrw2}{V. Barger, S. Geer, R. Raja, K. Whisnant, Phys.Rev. {\bf D62},
013004 (2000); Phys.Rev. {\bf D62}, 073002 (2000) }

%[41]
\bibitem{lb}{ I. Mocioiu, R. Shrock, A.I.P. Conf. Proc. Conf. Proc. 533 (2000);
I. Mocioiu, R. Shrock, Phys. Rev. {\bf D62}, 053017 (2000).}

%[42]
\bibitem{jeff} {J. Nelson, Private communication.}

%[43]
\bibitem{ray} {R. Stefanski, R. Bernstein, Private communication.}

%% additions from Debbies table email


%[29]
%\bibitem{e734result} L.A.Ahrens, {\em et al}, BNL-E734 Collaboration, 
%Phys. Rev. {\bf D31}  2732 (1985).  

%[31]
%\bibitem{bnle734} Bob Shrock will know whom to quote on this 


\end{thebibliography}
\end{document}